# Optimization of superconducting properties of F-doped SmFeAsO by cubic anvil high-pressure technique

Mohammad Azam[1], Tatiana Zajarniuk[2], Hiraku Ogino[3], Shiv J. Singh[1]*

[1]*Institute of High Pressure Physics (IHPP), Polish Academy of Sciences, Sokołowska 29/37, 01-142 Warsaw, Poland*

[2]*Institute of Physics, Polish Academy of Sciences, Aleja Lotników 32/46, 02-668 Warsaw, Poland*

[3]*Research Institute for Advanced Electronics and Photonics, National Institute of Advanced Industrial Science and Technology (AIST), Tsukuba, Ibaraki 305-8568, Japan*

*Corresponding author:

 Email: sjs@unipress.waw.pl

https://orcid.org/0000-0001-5769-1787

# Abstract

We optimize the synthesis conditions for $SmFeAsO_{0.80}F_{0.20}$ (Sm1111) bulks using a cubic-anvil high-pressure (CA-HP) apparatus through both *ex-situ* and *in-situ* processes, applying pressures of up to 4 GPa and heating temperatures of up to 1600°C. A comprehensive characterization has been performed, including structural, microstructural, transport, and magnetic measurements. Our findings indicate that a modest growth pressure of approximately 0.5 GPa is sufficient for the formation of the Sm1111 phase in the *ex-situ* process. In contrast, the *in-situ* process requires higher synthesis pressure (4 GPa) and temperature (1400 °C for 1 hour) to achieve the Sm1111 phase with enhanced superconducting properties. Notably, the optimized *in-situ* process significantly reduces the reaction time needed for the formation of the Sm1111 phase compared to conventional synthesis process at ambient pressure (CSP), leading to an increase in the transition temperature by 3 K and improvements in critical current density ($J_c$). Conversely, the optimized *ex-situ* process results in an onset transition temperature ($T_c$) of approximately 53 K, similar to that of CSP, though it enhances the $J_c$ by an order of magnitude. Despite these advancements, a small amount of impurity phases, as observed during CSP, persists in all Sm1111 samples prepared through either the *in-situ* or *ex-situ* CA-HP processes. These results suggest that the *in-situ* process under optimized conditions (1400 °C, 4 GPa for 1 hour) can effectively improve the superconducting properties of Sm1111. Additionally, a comprehensive analysis comparing these results with high gas pressure techniques, spark plasma sintering, and CSP methods suggests that a small amount of impurity phases in Sm1111 is persistent and cannot be completely eliminated by various pressure techniques, even at the applied pressure of up to 4 GPa. These findings could be beneficial for both fundamental and applied studies for the further progress of iron-based superconductors (FBS).

# Introduction

The second high-$T_c$ iron-based superconductor (FBS) has significant potential for practical applications due to its remarkable superconducting properties and is very rich from a chemistry point of view, where many kinds of doping are reported and provide a very rich phase diagram [1], [2], [3], [4] [5] [6]. Among these materials, F-doped SmFeAsO, belonging to the 1111 family (*RE*FeAsO, *RE* = Sm, La, Nd, Pr, etc.), provides the highest $T_c$ value of approximately 58 K , alongside large upper critical fields ($H_{c2}$~100 T) and a high critical current density ($J_c \approx 10^6$ A/cm²) [7] [8] [9] [10]. The characteristics of the 1111 family have huge potential for the practical applications but due to the presence of impurity phases in the prepared samples, attention of researcher diverted to other families of FBS and it is difficult in producing dense, high-quality bulk samples by conventional synthesis processes at ambient pressure (CSP). Conventional synthesis methods have difficulty dealing with the complex chemical and structural nature of FBS. The high volatility of arsenic and fluorine during synthesis often results in compositional deviations, while the formation of secondary phases and porosity in polycrystalline samples further degrades superconducting performance [11] [12, 4]. Moreover, achieving homogeneous doping levels in F-doped FBS like SmFeAsO$_{1-x}$F$_x$ remains challenging, as fluorine incorporation is sensitive to synthesis conditions. Over the past decade, efforts to enhance $T_c$ and $J_c$ of 1111 family through CSP method have almost stabilize with the maximum $T_c$ of 58 K [13, 14] [15] and low bulk $J_c$ (~$10^3$ A/cm²). These limitations suggest a need for alternative synthesis methods capable of overcoming the fundamental issues of sample quality, stoichiometry control, grain connectivity and improvement of their superconducting properties [11] [16]. In this direction, high-pressure synthesis has emerged as a transformative approach to address these challenges, enabling the growth of both single crystals and polycrystalline samples with superior structural and superconducting properties by applying pressures up to 6 GPa using solid-state medium or 1.8 GPa with high gas pressure techniques. High-pressure techniques diminish the volatility of constituent elements, enhance sample density, and improve grain connectivity. A milestone achievement in FBS research was the growth of the first 1111-family single crystal under high pressure by Karpenski et al. [17] [18], demonstrating the potential of this method to produce high-quality samples. Notably, Ren Zhi-An *et al.* reported that high-pressure synthesis at ~6 GPa enabled the growth of SmFeAsO$_{0.90}$F$_{0.10}$ polycrystals with $T_c$ values reaching 55 K [19] but had prepared two samples at the random pressure and heating temperature. Similarly, Zhi-An Ren *et. al.* synthesized NdFeAsO$_{0.89}$F$_{0.11}$ polycrystalline samples under 6 GPa with onset $T_c$ of 51.9 K [20]. Subsequent studies have confirmed that the synthesis of oxygen-deficient $RE$FeAsO$_{1-x}$, is only possible with high-pressure synthesis method [21, 22]. Recently, we reported the improved superconductivity in FeSe$_{0.5}$Te$_{0.5}$ [23] and CaKFe$_4$As$_4$ [24]using the high gas pressure and high temperature (HP-HTS) method, along with F-doped SmFeAsO. We demonstrated the enhanced sample quality and superconductivity, and recent reviews have further emphasized the effectiveness of high-pressure synthesis in optimizing the structural and superconducting properties of FBS. Deng *et al.* reported that high-temperature superconductivity induced in FeSe single crystals under pressure can be retained even after the pressure is removed at low temperature (~4 K), called the pressure quenching (PQ) process. This approach depicted the superconductivity up to 37 K at ambient pressure for these crystals for a few weeks, which is about four times higher than FeSe's transition temperature of approximately 9 K, achieved by the CSP method [25].

The multi anvil cell technique, particularly the cubic anvil high pressure (CA-HP) apparatus, is notable among various high-pressure techniques for its capability to deliver precise control over high pressure and temperature, potentially achieving levels above 5 GPa. This capability makes it particularly well-suited for synthesizing complex materials. The cubic anvil technique employs six anvils to apply uniform pressure from multiple directions, resulting in stable conditions and heating temperatures that can surpass 2000°C [26]. Additionally, its capacity to sustain high-pressure conditions over extended periods supports thorough reaction kinetics, thereby enhancing the production of phase-pure, dense samples with improved superconducting properties [12]. Very few studies have addressed the growth of fluorine-doped Sm1111 using this technique [21] [17], with only one sample prepared under arbitrary pressure and temperature conditions. However, a systematic investigation employing the CA-HP process is essential for this promising candidate to better understand its effects on both the sample quality and the superconducting properties of Sm1111. In light of this need, this study focuses particularly on the optimal fluorine doping levels for SmFeAsO, specifically targeting SmFeAsO$_{0.80}$F$_{0.20}$.

In this study, we have optimized the high-pressure synthesis of F-doped SmFeAsO bulk using the cubic anvil cell high pressure (CA-HP) technique, exploring both *ex-situ* and *in-situ* processes under various pressures (0.5-4 GPa), heating temperatures (900-1600°C), and durations (1-2 hours). Additionally, we have prepared SmFeAsO$_{0.80}$F$_{0.20}$ bulk using the CSP method to gain a clearer understanding of how the CA-HP process influences the formation of secondary phases and the superconducting properties of Sm1111. Extensive characterizations are conducted to address challenges related to sample homogeneity, grain connectivity, and fluorine incorporation through high-pressure growth, as well as to evaluate their impact on superconducting properties. Our findings suggest that the *in-situ* process enhances sample quality and superconducting properties more effectively, while the *ex-situ* process enhances the critical current density by keeping almost the same superconducting transition as observed by the CSP process. Furthermore, the results obtained are compared to those of $SmFeAsO_{0.80}F_{0.20}$ synthesized via the spark plasma sintering (SPS), and high gas pressure and high-temperature synthesis (HP-HTS) method, providing a broader perspective on these three high-pressure techniques in relation to the conventional synthesis process.

## Experimental Details

First, bulk $SmFeAsO_{0.80}F_{0.20}$ samples were synthesized using a single-step method through the conventional solid-state reaction process at ambient pressure (CSP). High-purity elemental precursors were utilized: samarium (Sm, 99.9%), iron (Fe, 99.99%), arsenic (As, 99.99%), iron oxide ($Fe_2O_3$, 99%), and iron fluoride ($FeF_2$, 99%). Due to the high reactivity of arsenic element, SmAs was synthesized first as a precursor. This compound was prepared by thoroughly mixing stoichiometric amounts of Sm and As, pelletizing the mixture, placing it in a tantalum (Ta) tube, and sealing this tube within an evacuated quartz tube. The sealed pellet was then heat-treated at 550 °C for 15 hours to form SmAs. In the subsequent step, the resulting F-doped Sm1111 was mixed with all remaining precursors according to the nominal stoichiometric formula of $SmFeAsO_{0.80}F_{0.20}$, with one batch weighing approximately 6-7 grams. The mixed powder was then pressed into 4-5 disk-shaped pellets (11 mm in diameter) using a hydraulic press at 200 bar. These pellets were enclosed in a Ta tube, which was sealed inside an evacuated quartz ampoule to prevent oxidation. The sealed ampoule was sintered at 900 °C for 45 hours [27]. All weighing, mixing, and pressing processes were conducted within a high-purity argon-filled glove box, ensuring that oxygen and moisture levels remained below 1 ppm to maintain the stability of the reactive components. The sample, prepared in a single step via CSP, is referred to as a parent P sample. We systematically characterized this sample using structural, microstructural, transport, and magnetic measurements to verify the formation of the samples and to compare our findings with previously reported data [27].

The *ex-situ* high-pressure synthesis of F-doped Sm1111 was conducted using a cubic anvil high-pressure (CA-HP) apparatus, as illustrated in the schematic diagram in Figure 1. The starting materials, primarily the parent sample, were encapsulated in a closed crucible made of boron nitride (BN) and then embedded in solid media (pyrophyllite [$Al_2Si_4O_{10}(OH)_2$]), which transmits pressures typically reaching several GPa to the samples [26]. A graphite tube with a cap served as the heater. The configuration of the sample cell is depicted in Figure 1(c). In the series of samples prepared for *ex-situ* high-pressure synthesis, each seal contained approximately 0.45 grams of the parent sample within the boron nitride crucible. For the *ex-situ* process, the parent P samples underwent high-pressure synthesis at 900 °C as the lowest suitable heating temperature for phase formation [23] for 1 hour under varying pressures ranging from 0.5 to 4 GPa. Conversely, in the *in-situ* process, arsenic (As, 99.99%), iron oxide ($Fe_2O_3$, 99%), iron fluoride ($FeF_2$, 99%), and prepared SmAs were utilized as initial precursors. The nominal stoichiometric formula of $SmFeAsO_{0.80}F_{0.20}$ was employed, weighing 2 grams and mixed thoroughly with a mortar and pestle before being filled into the cell of the CA-HP system, as shown in Figure 1(c). First, we have mixed the initial precursors according to the chemical formula $SmFeAsO_{0.8}F_{0.2}$ for a total of 3 grams materials. After grinding, approximately 0.45 grams of the initial material were used for each *in-situ* experiment. Since our initial *in-situ* synthesis experiments were conducted at the heating temperatures below 1100°C and at lower pressure (1-2 GPa), the Sm1111 phase did not form. Additionally, Ren Zhi-en et al. [21] reported the synthesis of an Sm1111 bulk sample at 6 GPa and 1250°C for 1 hour using an *in-situ* process in 2008, where only one or two samples were prepared. Based on these findings, we decided to carry out our *in-situ* synthesis process at a pressure of 4 GPa with a heating duration of 1 to 2 hours and heating temperatures ranging from 1100 to 1600°C. After the preparation of both *ex-situ* and *in-situ* high-pressure synthesized samples, various characterizations were performed, similar to those conducted on the parent P sample. The corresponding block diagrams for the *in-situ* and *ex-situ* processes are illustrated in Figure 2. All prepared samples are listed in Table 1 with the sample codes.

X-ray diffraction (XRD) patterns were collected using a Rigaku SmartLab 3 kW diffractometer to identify the presented phases. This system employed with filtered Cu-Kα radiation (λ =1.5418 Å) and operated at 30 mA and 40 kV. Diffraction patterns were recorded from 20° to 70° with a step size of 0.01 deg/min, using a Dtex250 linear detector. Lattice parameters were analyzed with Rigaku's PDXL software in conjunction with the ICDD PDF4+ 2025 database. With a four-probe measurement setup connected to a closed-cycle refrigerator (CCR), we assessed the temperature dependence of electrical resistivity measurements, which were performed from 7 K to 300 K, with the applied current varied from 5 mA to 20 mA. Magnetic properties were characterized using a vibrating sample magnetometer (VSM) integrated with a Physical Property Measurement System (PPMS). Under a static magnetic field of 20 Oe, magnetic susceptibility measurements were conducted over a temperature range of 5

K to 60 K. Additionally, a magnetic hysteresis loop curve (*M-H*) was acquired at 5 K over a range of 0 to 9 T of magnetic field.

## Result and analysis

### 1. XRD Analysis

To evaluate the purity and crystalline structure of the prepared $SmFeAsO_{0.80}F_{0.20}$ polycrystalline samples, we conducted powder X-ray diffraction (XRD) measurements, with the resulting patterns shown in Figure 3. These analyses included the parent material (P) and several Sm1111 samples synthesized using *ex-situ* and *in-situ* CA-HP processes. A tetragonal ZrCuSiAs-type structure (space group P4/nmm) [28] was identified as the main phase in all patterns, which confirms the phase formation for Sm1111 as reported in existing literature [29]. We have analyzed the phase purity and computed the lattice parameters, as listed in Table 1. The lattice parameters derived for the parent P sample are $a$ = 3.928(7) Å and $c$ = 8.497(9) Å, corroborating earlier findings [13, 27]. It is noteworthy that slight variations in the lattice parameters could fall within the error margin without showing any specific trend related to different pressure synthesis, likely due to the presence of secondary phases, as previously noted [30, 13]. Minor amounts of SmOF, SmAs, and FeAs phases were also identified as secondary phases, as depicted in Figure 3. Figure 3(a) displays the XRD pattern of the *ex-situ* samples, which include the parent $SmFeAsO_{0.80}F_{0.20}$ samples. As illustrated in the figure, most of the peaks for the samples correspond to the Sm1111 phase; however, a small quantity of the secondary phases, including SmOF, SmAs, and FeAs, is also evident. The parent P exhibits SmOF and SmAs as impurity phases, whereas the *ex-situ* processed samples contain a minor amount of the FeAs phase. With increasing pressure during the *ex-situ* process, the SmOF and SmAs phases remain nearly constant, while a slight enhancement is observed in the amount of the FeAs phase. Interestingly, the observed lattice parameter '*a*' for all *ex-situ* processed Sm1111 samples is nearly constant and comparable to that of the parent P, whereas a slight reduction is noted in the lattice parameter '*c*', indicating a slightly higher fluorine content within the lattice compared to the parent P. To clarify, we present the zoomed-in XRD patterns of the main peak (102) of the tetragonal superconducting phase of Sm1111 in Figure 3(b) for the *ex-situ* samples. The peak position for HP-1 sample prepared at 0.5 GPa is almost the same as the parent P, suggesting almost no reduction in fluorine content within the superconducting lattice. Conversely, the sample HP-2 prepared at 1 GPa shows a little peak shift to the higher angle, suggesting a little change in the fluorine content in the superconducting lattice. As pressure increases further, a little peak shift to the left (lower angle) becomes more pronounced, indicating that samples processed *ex-situ* at lower pressures (0.5 GPa) are relatively superior. Figure 3(c) displays the XRD patterns for Sm1111 samples prepared via *in- situ* processes alongside the parent compound. It appears that the SmOF impurity slightly increases in the *in-situ* processed Sm1111 samples, while the SmAs phase is slightly suppressed, as noted in Table 1. The sample HP-6 prepared at 1400 °C for 1 hour exhibits a lower amount of SmOF and a lower lattice parameter '*c*' compared to the HP-5 sample prepared at 1100°C. However, increasing the heating time to 2 hours (HP-8) leads to a resurgence of the SmOF phase along with an increase in lattice '*c*'. Figure 3(d) highlights the peak 102 area, where the sample prepared at 1100°C shows a peak position nearly identical to that of the parent P. In contrast, a peak shift to the higher angle is observed for HP-6 sample, suggesting an increased fluorine content within the superconducting lattice. Further increases in heating temperature to 1600°C or heating time to 2 hours appear to stabilize the peak position, keeping it nearly the same as that of the parent P. Consequently, the observed lattice '*c*' values in Table 1 for these samples are almost identical to those of the parent P. These findings from *in-situ* processed samples indicate that the synthesis conditions of 1400 °C for 1 hour are more favorable for enhancing the actual fluorine content in the F-doped Sm1111.

## 2. *Elemental Mapping and Analysis*

We have used energy-dispersive X-ray (EDAX) spectroscopy to perform elemental analysis and mapping on polished samples. The sample surfaces were prepared using progressively finer grades of sandpaper without any lubricants. This evaluate to assess how the constituent elements are distributed within the bulk samples. Selected results are shown in Figure 4, with additional data provided in Supplementary Figure S1. As seen in Figure 4(i), the parent P sample displays a nearly homogeneous composition, with Sm, Fe, As, O, and F distributed evenly across the mapped area. Nonetheless, certain localized spots show an enrichment in Sm, As, F, and O, which indicates the presence of minor SmAs and SmOF/$Sm_2O_3$ phases and confirms the XRD findings. The mapping for the *ex-situ* processed HP-1 sample, presented in Figure 4(ii), reveals a slightly less uniform distribution of Sm, Fe, As, O, and F in some areas when compared to the parent P. This observation is consistent with the XRD measurements that suggested the formation of secondary phases such as $Sm_2O_3$, SmAs, and FeAs. Notably, impurity phases are uniformly distributed throughout the parent P, but during the high-pressure growth process, these impurities tend to accumulate in specific areas of the sample, while the remainder appears nearly homogeneous. The areas of impurity are small in the HP-1 sample, but this trend increases with pressure in the HP-2 sample (Figure S1(i)). Upon further increasing the pressure to 4 GPa, these secondary phases began to redistribute within the samples, as observed in HP-3 and HP-4 (Figure S1(ii)-(iii)). which may pose challenges for critical current properties [31] [30] [32]. In these samples, the inhomogeneity of the constituent elements indicates the presence of impurity phases, as evidenced by the XRD patterns and detailed in Table 1. This *ex-situ* processed samples suggest that the HP-1 sample exhibits superior elemental mapping compared to the others.

The elemental mappings collected from the polished samples processed by the *in-situ* method are presented in Figure S1(iv)-(vi) and Figure 4(iii). The sample HP-5 exhibits significant inhomogeneity with elements such as Sm, Fe, As, and O/F, confirming the presence of SmOF, SmAs, and FeAs phases. When the heating temperature is increased to 1400 °C i.e., for sample HP-6, the resulting samples are nearly homogeneous; however, some impurity phases can still be observed. In contrast, the sample prepared at 1600°C displays greater inhomogeneity compared to HP-6, as illustrated in Figure S1(v). These analyses indicate that a heating temperature of 1400 °C is more favorable for the *in-situ* process of Sm1111 at 4 GPa. Consequently, one sample, HP-8, was prepared with a prolonged heating time of 2 hours at 1400 °C, and its elemental mapping is shown in Figure S1(vi). The elemental mapping reveals several inhomogeneous regions, suggesting the same types of impurities noted in other samples. Overall, these elemental mappings imply that the conditions used for sample HP-6 are more suitable for the *in-situ* process.

### 3. Microstructural analysis

Microstructural investigations were conducted on polished bulk samples of F-doped $SmFeAsO_{0.80}F_{0.20}$, which were synthesized by CSP at the ambient pressure (parent), *ex-situ*, and *in-situ* high-pressure techniques. The samples were prepared with progressively finer grades of emery paper and analyzed using a scanning electron microscope (SEM) in backscattered electron (BSE) mode at various magnifications. The BSE imaging technique is valuable as it simultaneously reveals both chemical contrast and microstructural features, allowing for precise identification of phase distribution and porosity. Figure 5(a-f) displays the representative BSE images for the selected samples: the parent P (ambient pressure), the optimized *ex-situ* sample HP-1 (0.5 GPa at 900 °C), and the optimized *in-situ* sample HP-6 (4 GPa at 1400 °C for 1 hour). In these images, three distinct contrasts are evident: bright regions correspond to impurity phases such as $Sm_2O_3$ or SmOF, light grey areas represent the F-doped SmFeAsO (Sm1111) superconducting phase, and black regions indicate pores with occasional contributions from minor phases like SmAs or FeAs. The parent P sample, illustrated in Figures 5(a-b), displays a microstructure characterized by fine grains interspersed with numerous nanopores. This indicates many weak intergranular connections [27] [33], as well as a low overall sample density. Additionally, impurity phases such as $Sm_2O_3$ and SmAs are distributed relatively uniformly throughout the sample, as supported by elemental mapping. In the *ex-situ* high-pressure sample HP-1 (Figures 5(c-d)), the Sm1111 phase appears more homogeneously distributed, accompanied by a notable reduction in porosity. However, a few localized regions of $Sm_2O_3$ and SmAs remain observable. Despite these impurities, the microstructure demonstrates the improved connectivity between superconducting grains. Importantly, all these *ex-situ* and *in-situ* processed sample exhibits a relative density of approximately 98%, based on the theoretical density of 7.1 g/cm³ for Sm1111 [5], representing a substantial improvement over the parent P sample (sample density~50%). Increasing the synthesis pressure beyond 0.5 GPa (e.g., at 1, 3, or 4 GPa) enhances the accumulation of secondary phases but does not result in further improvement of the sample, as illustrated in Figure S2(a)-(i). Instead, these samples show a more pronounced accumulation of impurity phases such as $Sm_2O_3$ and SmAs/FeAs, along with occasional agglomeration and microstructural inhomogeneity. These findings suggest that excessive pressure may negatively impact phase stability during *ex-situ* synthesis and support the conclusion that 0.5 GPa at 900 °C for 1 hour (HP-1) is the most favorable condition for the *ex-situ* high-pressure growth of F-doped SmFeAsO bulks, as reported in high gas pressure techniques [30]. The *in-situ* high-pressure samples are illustrated in Figures 5(e-f) and the supplementary Figure S2(j-r) for HP-5 to HP-8. The sample HP-6 displays a well-compacted microstructure with well-connected grain boundaries; however, minor regions of $Sm_2O_3$ and SmAs/FeAs impurities remain detectable. The sample HP-5, synthesized at a lower *in-situ* temperature (e.g., 1100 °C), exhibits a more incoherent microstructure with reduced porosity. In contrast, the sample HP-7, sintered at 1600 °C for 1 hour, along with the extended-duration sample HP-8, processed at 1400 °C for 2 hours (shown in Supplementary Figure S2), demonstrates increased porosity and impurity segregation. This is likely attributed to excessive grain growth or fluorine loss at elevated temperatures and extended processing times. Overall, the microstructural analysis indicates that *ex-situ* synthesis at 0.5 GPa and 900 °C for 1 hour yields the most uniform and densified microstructure among all *ex-situ* batches, without necessitating high-pressure conditions exceeding 1 GPa. Conversely, *in-situ* synthesis at 4 GPa and 1400 °C for 1 hour (HP-6) achieves the most optimized grain connections and phase purity compared to other pressure conditions. These microstructural findings align well with the observed trends in the structural and elemental mapping measurements discussed earlier, underscoring the significant impact of controlled pressure and temperature on the quality of SmFeAsO$_{0.80}$F$_{0.20}$ bulk superconductors.

### 4. Transport properties

The temperature-dependent resistivity of SmFeAsO$_{0.80}$F$_{0.20}$ bulk, prepared via an *ex-situ* process under various high-pressure conditions (ranging from 0.5 to 4 GPa), is illustrated in Figure 6(a) for temperatures between 7 K and 300 K. The parent P sample, prepared at ambient pressure, shows a room temperature resistivity of approximately 5.5 mΩ-cm. This resistivity decreases linearly with cooling and indicates a superconducting onset transition at 53.1 K. This behavior aligns with previous reports on F-doped Sm1111 superconductors [27] [34] [35]. When the parent P sample is reground and subjected to an *ex-situ* sintering process at 0.5 GPa and 900 °C for 1 hour (i.e., HP-1), a significant reduction in resistivity is observed, approximately five times lower than that of the parent P. This improvement is attributed to enhanced sample density and grain connectivity, as confirmed by microstructural observations. A further increase in the applied pressure to 1 GPa (i.e., HP-2) maintains a nearly identical resistivity behavior and exhibits a slight decrease compared to HP-1, indicating a moderate enhancement in grain connectivity. However, when the synthesis pressure is increased to 3 GPa (HP-3) and 4 GPa (HP-4), there is a slight rise in resistivity across the entire temperature range. This trend correlates with increased porosity and the accumulation of impurity phases, such as Sm$_2$O$_3$, as observed from the microstructural analysis. These findings suggest that while moderate pressure improves the intergrain connectivity of the samples, excessive pressure (greater than 1 GPa) adversely affects it due to the introduction of secondary phases between the grains. To better understand the superconducting transitions, Figure 6(b) illustrates the low-temperature resistivity (40-60 K) of the *ex-situ* processed samples. The Parent P sample displays an onset transition temperature ($T_c^{onset}$) of 53.1 K, while the HP-1 sample shows a slightly reduced $T_c^{onset}$ of approximately 53 K, likely due to slightly improved or almost the same fluorine incorporation compared to the sample P, as indicated by XRD analysis. The way for obtaining the onset transition temperature ($T_c^{onset}$) and offset transition temperature ($T_c^{offset}$) of the parent P is shown in supplementary figure S3, and a similar approach is used for the other samples. When the synthesis pressure is increased to 1 GPa, the $T_c^{onset}$ is 52.6 K, whereas HP-3 and HP-4, synthesized at 3 GPa and 4 GPa, respectively, demonstrate nearly identical $T_c$ values (~52.5 K); however, their transition width is broader than that of the parent P sample, suggesting reduced homogeneity or issues with intergrain connections. The widening of the transition width with increased pressure reflects a balance between enhanced grain connectivity in specific regions and the persistence of impurity phases, which diminish the improvements in electrical performance [30]. These findings support the conclusion that a pressure range of 0.5 GPa at 900 °C for one hour represents the most favorable *ex-situ* growth conditions for the formation of the superconducting 1111 phase.

The *in-situ* synthesized samples (HP-5 to HP-8), prepared under a constant pressure of 4 GPa at varying temperatures and durations, had their resistivity graphs with temperature variation presented in Figure 6(c) up to room temperature. All samples exhibited metallic behavior, characterized by a nearly linear decrease in resistivity leading to the superconducting transition. When compared to the parent P, the resistivity values of the samples were significantly lower across the entire temperature range, indicating improved density and grain connectivity. The low-temperature resistivity profiles, illustrated in Figure 6(d), revealed onset $T_c$ values ranging from 52.7 to 56 K. The $T_c^{onset}$ increased for HP-5 and HP-6 compared to the parent P, while further increases in heating temperature and duration led to a reduction in the transition temperature. Among these samples, HP-6 demonstrated the highest $T_c^{onset}$ at approximately 56 K, which can be attributed to enhanced fluorine substitution in the crystal lattice, as confirmed by structural analysis. The low-temperature behavior of these samples is illustrated in Figure 6(d). The HP-5 sample exhibits a very broad transition with $T_c^{onset}$~54.8 K, while the HP-6 sample, prepared at 1400 °C for 1 hour, demonstrates improved grain connectivity with $T_c^{onset}$~56 K and a reduced transition width. An increase in heating temperature and duration leads to broader transitions, indicating the presence of secondary phases and sample inhomogeneity, as confirmed by structural and microstructural analyses. The $T_c^{onset}$ value is observed at 53.9 K and 52.7 K for HP-7 and HP-8, respectively. The transition width ($\Delta T$) for the HP-6 sample remains narrow (~2 K), suggesting relatively uniform superconducting properties. In contrast, other *in-situ* samples show slightly broader transitions, which indicate weaker grain connectivity, likely due to residual $Sm_2O_3$ and SmAs phases, as supported by microstructural and XRD analyses. Among the *in-situ* samples, the one synthesized at 4 GPa and 1400 °C for 1 hour, namely HP-6, exhibits reduced impurity areas and improved phase purity, correlating with enhanced grain connectivity. These findings suggest that these conditions represent the optimal synthesis parameters for the *in-situ* synthesis process of Sm1111, resulting in high $T_c$, a dense microstructure, and phase uniformity. In summary, the resistivity analysis indicates that *ex-situ* synthesis at 0.5 GPa and 900 °C achieves the optimal combination of low resistivity, a sharp superconducting transition, and minimal impurity phases. Conversely, the *in-situ* synthesis process at 4 GPa and 1400 °C for 1 hour demonstrates superior superconducting performance compared to other samples produced through *in-situ* synthesis, primarily due to enhanced fluorine incorporation and improved structural uniformity.

## 5. Magnetic properties

To confirm the superconducting behaviors and verify the Meissner effect, DC magnetic susceptibility measurements were performed in both zero-field-cooled (ZFC) and field-cooled (FC) modes across a temperature range of 5 to 60 K, with an applied field of 20 Oe. All samples, including the parent P, *ex-situ*, and *in-situ* high-pressure synthesized batches, were characterized. The magnetization values were normalized at 5 K to facilitate direct comparison. Figure 7(a) presents the normalized magnetization curves for the parent sample, as well as HP-1, HP-2, HP-3, and HP-4 samples. Each of these samples exhibits clear diamagnetic transitions, and the negative value of the FC signal indicates strong flux pinning and effective vortex trapping within the respective samples. The parent sample P demonstrate onset superconducting transitions near 52.5 K. In contrast, the HP-1 sample, prepared at 0.5 GPa, shows an almost same critical temperature ($T_c$) of 52.6 K, which is slightly lower and consistent with its resistivity behavior. Increasing the growth pressure to 1 GPa i.e., HP-2 has almost the same $T_c$ as that of HP-1, whereas the sample HP-3 and HP-4 have the onset $T_c$ of ~52 K. A notable two-step transition is observed for the parent P, reflecting weak intergranular coupling, which is a common characteristic of polycrystalline Sm1111 [36] [37] [32]. In contrast, all *ex-situ* high-pressure samples exhibit single-step transitions, indicating improved grain connectivity due to pressure-assisted densification. Figure 7(b) presents the normalized magnetization for *in-situ* samples synthesized at 4 GPa under varying temperatures and durations: HP-5, HP-6, HP-7, and HP-8. The HP-5 sample shows a critical temperature ($T_c$) of 54.1 K but has a very broad transition, as depicted in Figure 6(d), suggesting inhomogeneity within this sample. The HP-6 sample, sintered at 1400 °C for 1 hour, shows the highest $T_c^{onset}$ (~56 K), in agreement with transport measurements, which suggests slightly enhanced fluorine incorporation. Increasing the heating temperature to 1600 °C for the HP-7 sample reduces the $T_c$ value to 52.5 K, matching that of the parent P sample. Furthermore, increasing the heating time to 2 hours at 1400 °C for the HP-8 sample also leads to a more rapid reduction in $T_c$ (~51.4 K) compared to HP-7, indicating that the longer sintering time does not support the superconductivity, as also observed in the previously discussed resistivity measurements. Interestingly, the HP-7 and HP-8 samples exhibit almost a single-step transition, while HP-6 demonstrates a higher transition accompanied by an appearance of two-step transitions, similar to the parent P sample. These observations suggest that further optimization of the HP-6 sample could enhance the sharpness of the transition temperature and improve intergrain connectivity through slow cooling or warming.

For practical superconducting applications, high critical current density ($J_c$) is a crucial parameter. Magnetic hysteresis (*M-H*) loops were recorded at 5 K under magnetic fields up to 9 T for various samples, in which the *M-H* loop for the parent P, HP-1, and HP-6 samples are illustrated in the supplementary Figure S4. The hysteresis width (*ΔM*) was utilised to calculate the critical current density $J_c$ for all samples using the extended Bean model: $J_c = 20Δm/Va(1-a/3b)$, where $a$ and $b$ represent the sample dimensions, and $V$ is the volume. The calculated $J_c$ at 5 K for the *ex-situ* samples is presented in Figure 7(c) alongside the parent P sample. The parent P sample exhibited a maximum $J_c$ of approximately $10^3$ A/cm² at low fields, which gradually decreased with increasing magnetic field, consistent with previous studies on F-doped SmFeAsO [38]. The HP-1 sample demonstrated a significant improvement in $J_c$, reaching up to $9 \times 10^3$ A/cm² across the entire magnetic field range. In contrast, the sample processed at 1 GPa, designated as HP-2, showed a slight reduction in $J_c$ compared to HP-1, although it remained higher than that of the parent P. An additional increase in pressure led to a further decrease in $J_c$, approaching the values of the parent P sample, as observed in the HP-3 and HP-4 samples. This decline is attributed to the presence of impurity phases both between grains and within the grains, as noted in the microstructural analysis.

Figure 7(d) illustrates the magnetic field dependence of the critical current density ($J_c$) for samples processed using the *in-situ* method. The sample HP-5, synthesized at 1100°C, exhibits a lower $J_c$ value than the parent sample P, which is expected due to the presence of multiple impurity phases, as previously discussed. The HP-6 sample, prepared at 1400 °C for 1 hour, and the HP-7 sample, prepared at 1600°C for 1 hour, show nearly identical $J_c$ values across the entire magnetic field. However, extending the heating duration to 2 hours results in reduced $J_c$ values, as observed in the HP-8 samples. These *in-situ* processed samples indicate that the growth conditions for HP-6 and HP-7 yield the highest $J_c$, approximately $5 \times 10^3$ A/cm², surpassing the parent P sample and being slightly lower than the best *ex-situ* sample, HP-1. These enhancements can be attributed to higher density, improved grain connectivity, and better phase homogeneity, as demonstrated in previous structural and microstructural analyses. Notably, high-pressure samples show nearly double the sample density of the parent sample, which significantly contributes to the improvement in $J_c$. However, the presence of impurity phases, such as $Sm_2O_3$ and SmAs, poses challenges in achieving high $J_c$ values comparable to those reported for single crystal 1111. To further assess the superconducting $J_c$ performance, we calculated the flux pinning force ($F_p$) using the relation $F_p = \mu_0 H \times J_c$ [39]. The field dependence of $F_p$, presented in Supplementary Figure S5, confirms an enhanced pinning force resulting from the high-pressure synthesis process when compared to the parent sample. This improvement is consistent with previous findings and supports the conclusions drawn from the reported HP-HTS process of Sm1111 [30], where increased flux pinning contributes to the observed enhancement in $J_c$ for both *ex-situ* and *in-situ* processed samples.

### 6. Discussion

To thoroughly assess the effects of high-pressure synthesis using the cubic anvil cell technique, we extracted and analyzed the transport parameters for both *ex-situ* and *in-situ* synthesized SmFeAsO$_{0.80}$F$_{0.20}$ bulks. The parameters examined include the onset transition temperature ($T_c^{onset}$), transition width ($\Delta T$), room-temperature resistivity ($\rho_{300K}$), Residual Resistivity Ratio ($RRR = \rho_{300K} / \rho_{60K}$), and critical current density ($J_c$), as summarized in Figures 8 and 9. For the *ex-situ* synthesized samples, the onset transition temperature is observed at 53 K (HP-1 sample) which is slightly reduced to 52.6 K for HP-2 prepared at 1 GPa, followed by a near-saturation $T_c^{onset}$ for HP-4 and HP-5 samples synthesized at 3 GPa and 4 GPa, as shown in Figure 8(a). The transition width $\Delta T$, shown in Figure 8(b), is approximately 6 K for the HP-1 sample. In contrast, this value is slightly higher for the HP-2 sample prepared at 1 GPa compared to HP-1. The elevated pressure during synthesis has significantly broadened the transition temperature for the HP-3 sample (Figure 8(b)). This broadening remains nearly constant for the HP-4 sample, which was grown under a pressure of 4 GPa. The slight increase in $\Delta T$ at higher pressures indicates a potential degradation in sample homogeneity and grain connectivity. The room-temperature resistivity $\rho_{300K}$, illustrated in Figure 8(c), is around 1 mohm-cm for the HP-1 sample and slightly decreased for HP-2, which confirms its enhanced density and compactness. With increasing pressure, the resistivity at 300 K ($\rho_{300K}$) exhibits a slight upward trend, likely due to growing microstructural inhomogeneities and the accumulation of impurity phases. The calculated residual resistance ratio (*RRR*), illustrated in Figure 8(d), remains almost constant for the HP-1 and HP-2 samples prepared at low pressures up to 1 GPa. In contrast, it decreases linearly for the HP-3 and HP-4 samples prepared at higher pressures up to 4 GPa. This suggests that low-pressure synthesis has better sample homogeneity of the samples and intergranular connectivity. The performance of the magnetic critical current density, depicted in Figure 8(e), indicates that the HP-1 sample achieves the highest critical current density ($J_c$), approximately $9 \times 10^3$ A/cm² at 0.2 T and 5 K, which is nearly an order of magnitude greater than that of the parent sample synthesized using the CSP method. The $J_c$ value is slightly reduced for the HP-2 samples, while the higher-pressure samples (3-4 GPa) result in a further decrease in $J_c$. This reduction may be attributed to the increased presence of impurity phases, such as $Sm_2O_3$, across larger sample regions, as confirmed by microstructural analysis. These findings demonstrate that moderate pressure (0.5 GPa) and optimal sintering conditions (900 °C for 1 hour) are better conditions for the synthesis process of F doped Sm1111 through *ex-situ* CA-HP process.

The *in-situ* synthesised samples HP-5 to HP-8 were synthesized under a constant pressure of 4 GPa at varying temperatures and durations. As shown in Figure 9(a), the Parent P sample has an onset $T_c$ of 53 K, while HP-5 sintered at 1100 °C (1 h) and HP-6 sintered at 1400 °C (1 h) exhibit increased $T_c$ values of 54.8 K and 56 K, respectively. This demonstrates a ~3 K improvement over the CSP-processed Parent P, attributable to enhanced fluorine incorporation. Further increase of the synthesis pressure reduces the critical transition temperature. The transition width $\Delta T$, presented in Figure 9(b), has value around 6.4 K for parent P and 11 K for HP-5 sample. It reaches to the minimum value ~2.5 K for HP-6 sample processed at 1400 °C (1 h), suggesting better intergrain connectivity. Higher synthesis temperature or longer heating time further increase the transition broadening slightly, suggesting weak link grain boundaries. The room-temperature resistivity ($\rho_{300K}$), illustrated in Figure 9(c), decreases substantially in all *in-situ* samples, ranging from 1-2 mΩ·cm, compared to 5.5 mΩ·cm for the Parent P. This decrease is attributed to higher sample density and more compact grain structures. The *RRR* value, shown in Figure 9(d), is lower for HP-5 and reach to maximum for HP-6 heated at 1400 °C (1 h) sample, indicating better homogeneity. Further increase of heating temperature reduce *RRR* value. The corresponding critical current density $J_c$ of these samples at 0.2 T and 8 T are depicted in Figure 9(e). HP-5 has almost same $J_c$ at low field but at high field, the value reduced rapidly compare to that of the parent P sample. HP-6 and HP-7 have almost the same $J_c$ value at low and high magnetic field which exhibit the highest $J_c$ values, approximately $5 \times 10^3$ A/cm². But the longer heating time i.e., HP-8 shows the lower $J_c$ value than that of HP-6 sample. Among all these samples, the *ex-situ* synthesised HP-1 sample at 0.5 GPa and 900 °C for 1 hour exhibits the enhanced superconducting properties and sample quality, particularly in terms of critical current density ($J_c$) having almost constant $T_c$~53 K. This indicates that moderate-pressure 0.5 GPa under optimized thermal conditions significantly improves grain connectivity, phase uniformity, and flux pinning through *ex-situ* process. On the other side, the *in-situ* HP-6 sample synthesised at 4 GPa and 1400 °C for 1 hour demonstrates the highest superconducting transition temperature ($T_c$) of approximately 56 K, suggesting that high-pressure *in-situ* growth is more effective in achieving optimal fluorine incorporation and structural ordering compared to *ex-situ* process and also other *in-situ* processed Sm1111 samples.

Figure 10 shows a comparative overview of the magnetic field dependence of the critical current density, $J_c$, at 5 K for SmFeAsO$_{0.80}$F$_{0.20}$ synthesised using various synthesis processes: CSP (parent sample), and several pressure-based techniques: *Ex-situ* CA-HP process, *In-situ* CA-HP process, Spark Plasma Sintering (SPS), and [40] high gas pressure and high temperature techniques (HIP) [30]. The obtained $T_c^{onset}$ of the optimized F-doped SmFeAsO (SmFeAsO$_{0.80}$F$_{0.20}$) by these techniques are listed in Table 2. Interestingly, the onset $T_c$ of Sm1111 is almost same for HP-HTS, SPS and *ex-situ* CA-HP processes as that of CSP method, whereas the onset $T_c$ is increased by 3 K for the *in-situ* CA-HP process. Among these, the Parent P sample exhibited the lowest $J_c$, likely due to its low relative density (~50%) and the presence of impurity phases such as SmOF, SmAs. The limited current-carrying capabilities are attributed to poor grain connectivity, weak flux pinning, and non-superconducting inclusions during CSP process. *Ex-situ* HP sample, synthesized by sintering the precursor under 0.5 GPa at 900°C for 1 hour, demonstrated the highest critical current density among the bulk samples, reaching values close to ~$10^4$ at 0.2 T and 5 K. This significant enhancement is attributed to improved microstructure, possible better grain alignment, reduced porosity, refined grain boundaries, and improved pinning centre. In comparison, *In-situ* HP sample, synthesised at a pressure of 4 GPa and 1600°C for 1 hour, did not reach the $J_c$ value of the e*x-situ* HP sample. This suggests that one-step synthesis can complete the chemical reaction of Sm1111 phase formation but it is not sufficient to improve $J_c$ performance, possibly due to weak grain [27] [41] connection problem as also reported for CaKFe$_4$As$_4$ bulk [24]. The SPS and HIP samples show slightly improved $J_c$ as compared to the Parent P and fall between the sample processed by HP-Synthesis and CSP samples. SPS employs pulsed DC current and moderate uniaxial pressure (45 MPa) to achieve rapid densification and good grain connectivity [40]. This method nearly doubles the sample Sm1111 density compared to conventional solid-state processing (CSP) (from ~50% to ~98%), but the amount and nature of impurity phases (e.g., SmOF and SmAs) remain almost same as that of parent sample [30]. Due to this, a limited improvement in $J_c$ is observed during SPS process of F-doped SmFeAsO. Similarly, the HIP sample prepared by HP-HTS technique at 0.5 GPa has a slight improvement in the sample compaction (~3-4%) but have almost the same amount of impurity phase as that of CSP and SPS process. Interestingly, these impurity phases are not reduced even by this CA-HP method either *ex-situ* or *in-situ* processes; however, the onset transition and the critical current density are enhanced. Based on these findings, we need more optimization processes for the high-pressure growth techniques to completly reduce the impurity phases in F-doped Sm1111. The synthesis process and pressure optimisation can play an important role for density, better grain connectivity, as well as high superconductivity of F-doped Sm1111. Among the methods evaluated, *ex-situ* CA-HP synthesis under moderate pressure and controlled temperature is the most effective in enhancing the critical current density, whereas the *in-situ* process is able to enhance the superconducting transition by 3 K at high pressure (4 GPa) and high heating temperature (1400°C).

## Conclusion

The synthesis of SmFeAsO$_{0.80}$F$_{0.20}$ bulk superconductors was successfully optimized using a CA-HP technique through both *ex-situ* and *in-situ* processing routes, and the growth pressure effects on the superconducting properties of Sm1111 were systematically explored. The *ex-situ* method involved sintering conditions at 0.5-4 GPa and 900 °C for 1 h, while the *in-situ* synthesis was conducted at pressures up to 4 GPa and temperatures ranging from 1100°C to 1600 °C for 1 to 2 h. Comprehensive structural, microstructural, transport, and magnetic measurements confirmed that high-pressure synthesis significantly improved the sample density and superconducting properties for the optimized growth conditions. Sm1111 bulks prepared via the conventional solid-state reaction method exhibited a $T_c$ around 53.1 K, a lower sample density (~50%) and a $J_c$ of ~$10^3$ A/cm$^2$. Importantly, our findings indicate that *ex-situ* synthesis at the optimized conditions (900°C, 1 h at 0.5 GPa) is more effective for improving $J_c$ by one order of magnitude, having a constant $T_c$, while *in-situ* synthesis at higher pressures (4 GPa) and elevated temperatures (1400 °C, 1 h) enables the enhancement of $T_c$ by 3 K and good $J_c$ retention compared to the parent Sm1111. These enhancements are primarily attributed to improved flux pinning, higher grain connectivity, and densification. However, the amount and type of secondary phases, as observed in the parent Sm1111 by CSP, was not reduced even by CA-HP synthesis processes, which are similar to the previous reports based on HP-HTS and SPS processes. Our analysis suggests that the presence of small amounts of secondary phases during the phase formation of Sm1111 are not easy to eliminate even by applying different pressure techniques and are the main obstacle to achieve the high superconducting properties in these F-doped Sm1111. We conclude that high-pressure synthesis is a powerful method to enhance the superconducting properties of F-doped Sm1111 bulks; however, further research is still needed in this direction.

**Acknowledgments:**

The work was funded by SONATA-BIS 11 project (Registration number: 2021/42/E/ST5/00262) sponsored by National Science Centre (NCN), Poland. SJS acknowledges financial support from National Science Centre (NCN), Poland through research Project number: 2021/42/E/ST5/00262. We would like to acknowledge H. Kito, AIST for his assistance during the experiments with cubic anvil high-pressure (CA-HP) techniques.

**Table: :1** Sample codes, sample synthesis conditions, lattice parameters, the amount of impurity phases of the parent and other samples prepared by *ex-situ* and *in-situ* CA-HP processes.

| | Samples Code | Synthesis Condition | Lattice 'a' (Å) | Lattice 'c' (Å) | SmOF (%) | SmAs (%) | FeAs (%) |
|---|---|---|---|---|---|---|---|
| | P | 900 °C, 45 h, 0 GPa | 3.928(7) | 8.497(9) | ~7 | ~2 | ~1 |
| *Ex-situ* | HP-1 | 900 °C, 1 h, 0.5GPa | 3.927(2) | 8.476(4) | 8 | 3 | 2 |
| | HP-2 | 900 °C, 1 h, 1 GPa | 3.927(2) | 8.491(4) | 8 | 2 | 4 |
| | HP-3 | 900 °C, 1 h, 3 GPa | 3.928(2) | 8.473(3) | 10 | 3 | 5 |
| | HP-4 | 900 °C, 1 h, 4 GPa | 3.929(2) | 8.491(4) | 9 | 3 | 4 |
| *In-situ* | HP-5 | 1100 °C, 1 h, 4 GPa | 3.925(2) | 8.465 (5) | 14 | ~3 | ~2 |
| | HP-6 | 1400 °C, 1 h, 4 GPa | 3.921(2) | 8.457 (5) | 11 | ~1 | ~1 |
| | HP-7 | 1600 °C, 1 h, 4 GPa | 3.920(2) | 8.458 (5) | ~11 | 2 | 2 |
| | HP-8 | 1400 °C, 2 h, 4 GPa | 3.927 (2) | 8.478 (5) | 14 | ~1 | 2 |

**Table: 2** The onset transition temperature ($T_c^{onset}$) for SmFeAsO$_{0.8}$F$_{0.2}$ samples prepared using the conventional synthesis process at ambient pressure (CSP), and by various high pressure methods such as HP-HTS (sample noted as "HIP"), spark plasma sintering (sample noted as "SPS"), and cubic anvil high-pressure technique (CA-HP) under *ex-situ* process (sample noted as "*Ex-situ* HP") and *in-situ* process (sample noted as "*In-situ* HP").

| Samples | $T_c^{onset}$ |
|---|---|
| CSP | 53.1 |
| HIP | 53 [30] |
| SPS | 53.5 [40] |
| *Ex-situ* HP | 53 |
| *In-situ* HP | 56 |

**Figure 1**: Schematic illustration of **(a)** Typical heating temperature-time profile of the polycrystal growth of SmFeAsO$_{0.8}$F$_{0.2}$ **(b)** the high-pressure cubic anvil apparatus, and **(c)** the sample cell assembly, including pyrophyllite cube, graphite sleeve, boron nitride (BN) sample crucible, pyrophyllite pellets, graphite disks.

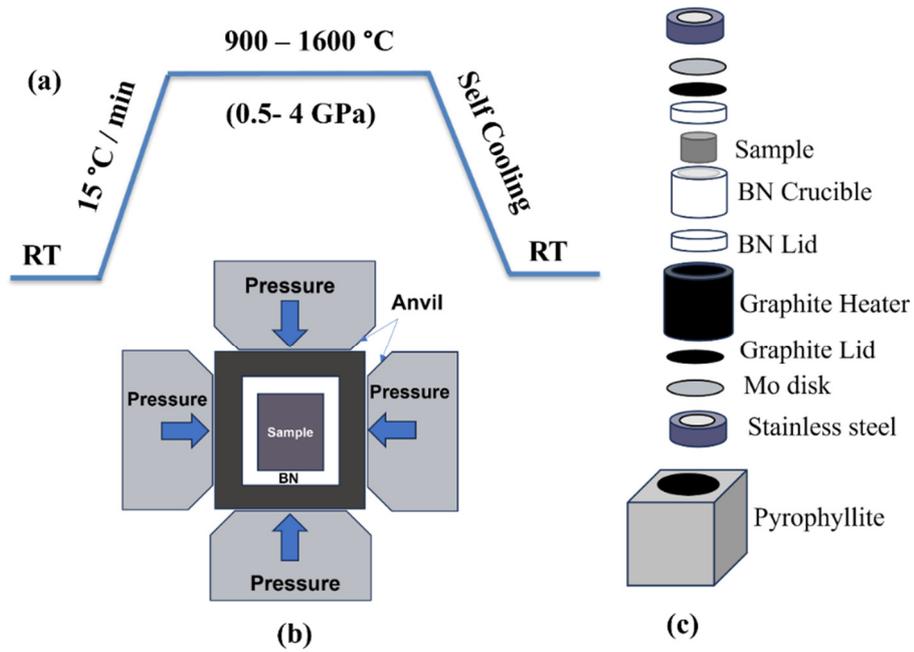

**Figure 2** Flow chart for the preparation of SmFeAsO$_{0.8}$F$_{0.2}$ bulks by CSP and CA-HP (*ex-situ* and *in-situ*) methods.

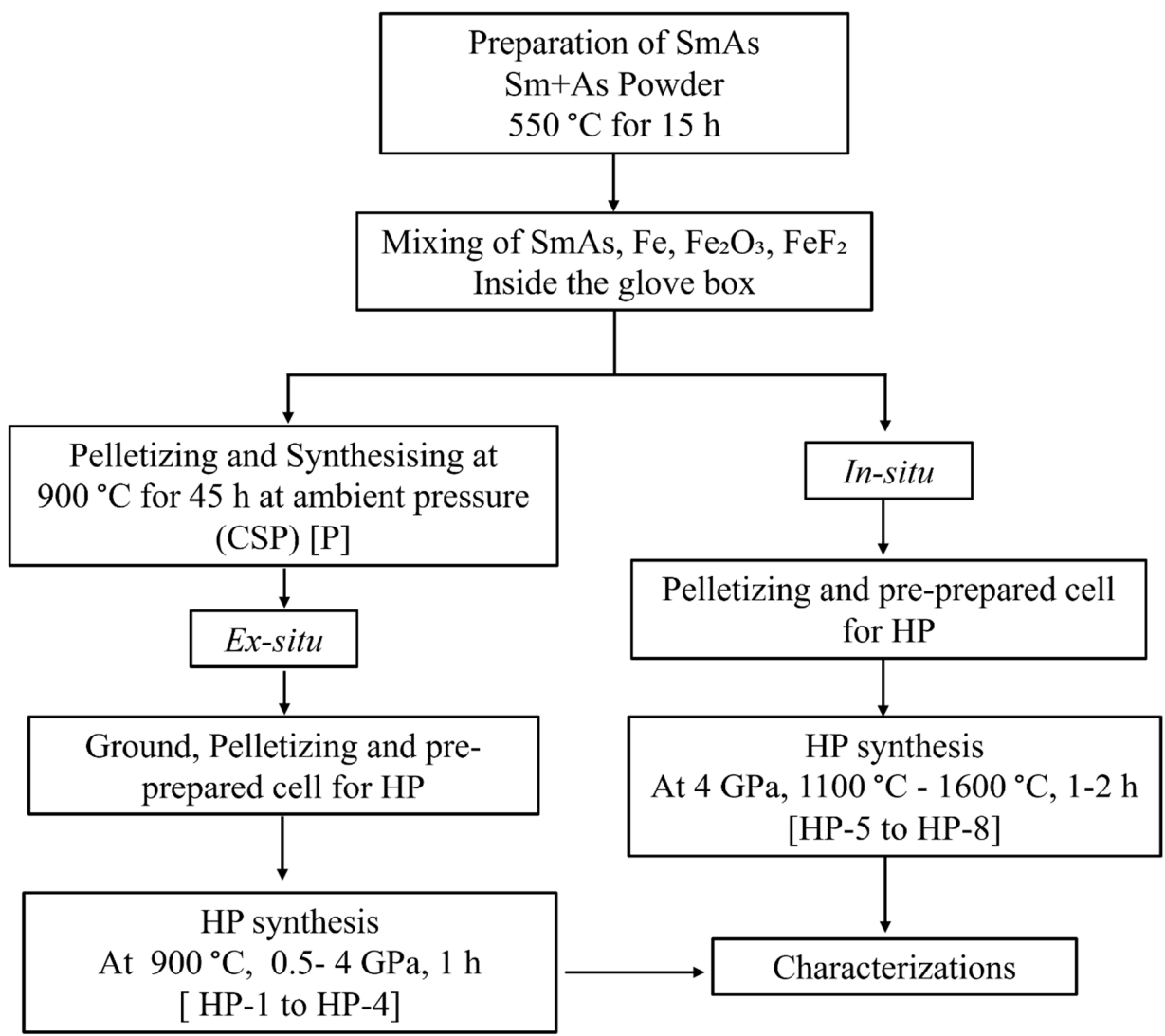

**Figure 3: (a)** Powder X-ray diffraction (XRD) patterns and **(b)** An enlarged view of the main (102) diffraction peak for SmFeAsO$_{0.80}$F$_{0.20}$ samples prepared by the CSP and CA-HP *ex-situ* process under different pressures. **(d)** Powder X-ray diffraction (XRD) patterns and **(e)** An enlarged view of the main (102) diffraction peak for SmFeAsO$_{0.80}$F$_{0.20}$ samples prepared by the CSP and CA-HP *in-situ* process under different heating temperatures at the applied pressure of 4 GPa.

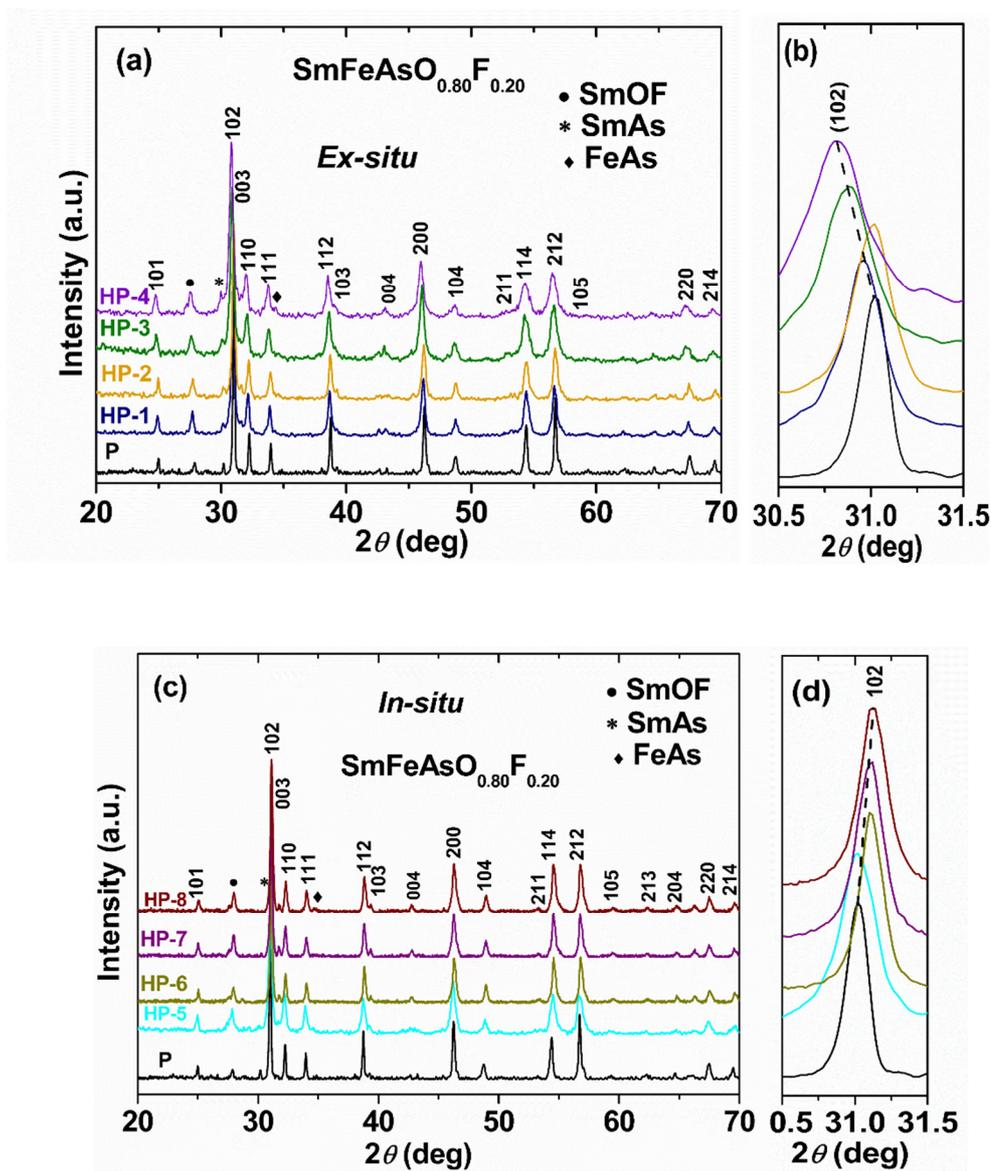

**Figure 4: (i)** Elemental mapping for all constituent elements of **(i)** Parent P prepared by CSP, **(ii)** HP-1 prepared by CA-HP at 900 °C, 0.5 GPa using an *ex-situ* process, and **(iii)** HP-6 prepared by CA-HP at 4 GPa for 1 h at 1400 °C using an *in-situ* process. The first and last images for each sample are SEM images and a combined image of all the constituent elements, respectively. The rest of the images depict the elemental mapping of individual elements Sm, Fe, As, O, and F.

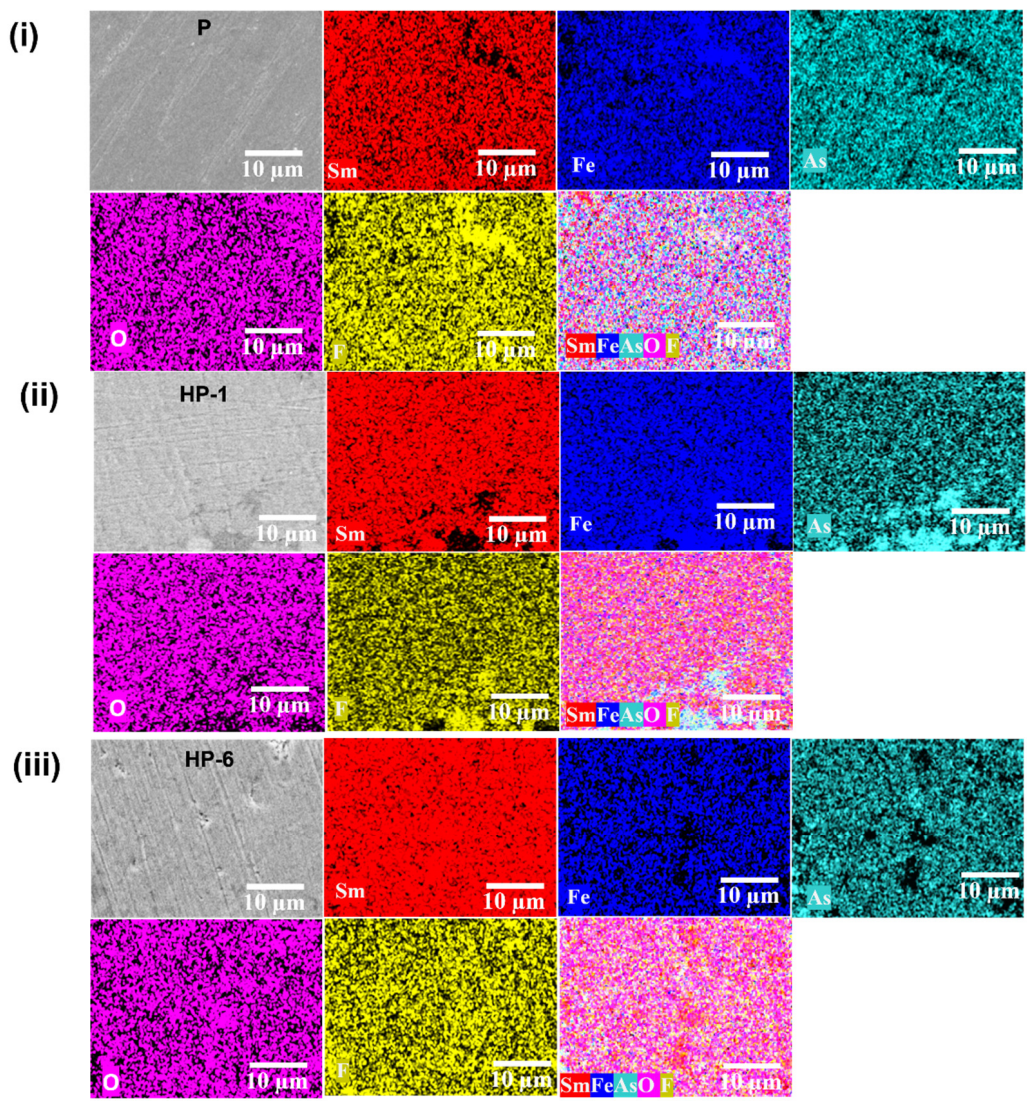

**Figure 5:** Backscattered electron images (BSE; AsB) of **(a)-(b)** Parent P, **(c)-(d)** HP-1, and, **(e)-(f)** HP-6. Bright contrast, light grey, and black contrast correspond to $Sm_2O_3$ (SmOF), the $SmFeAsO_{0.80}F_{0.20}$ phase and pores, respectively. One can note that black contrast can occasionally be SmAs/FeAs.

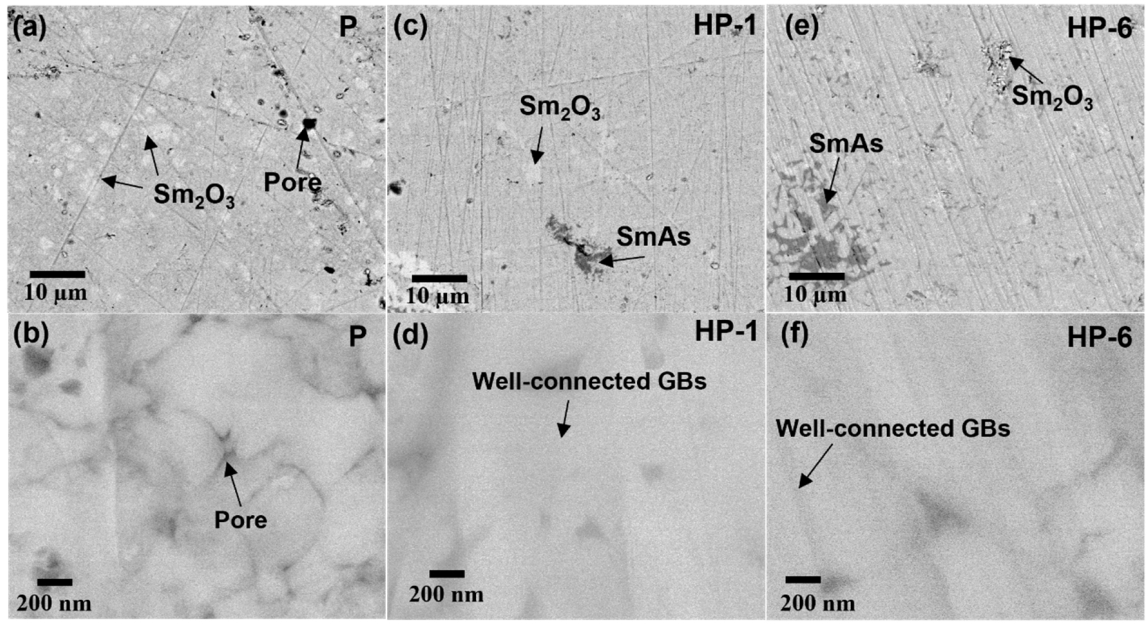

**Figure 6: (a)** The variation of normal state resistivity ($\rho$) with temperature up to room temperature **(b)** Low-temperature variation of the resistivity up to 60 K for various Sm1111 bulks by *ex-situ* CA-HP processes. The inset image of figure (a) depicts the temperature dependence of the resistivity of the parent P sample up to the room temperature. **(c)** The variation of normal state resistivity ($\rho$) with temperature up to room temperature **(d)** Low-temperature variation of the resistivity up to 60 K for various Sm1111 bulks prepared by the *in-situ* CA-HP process.

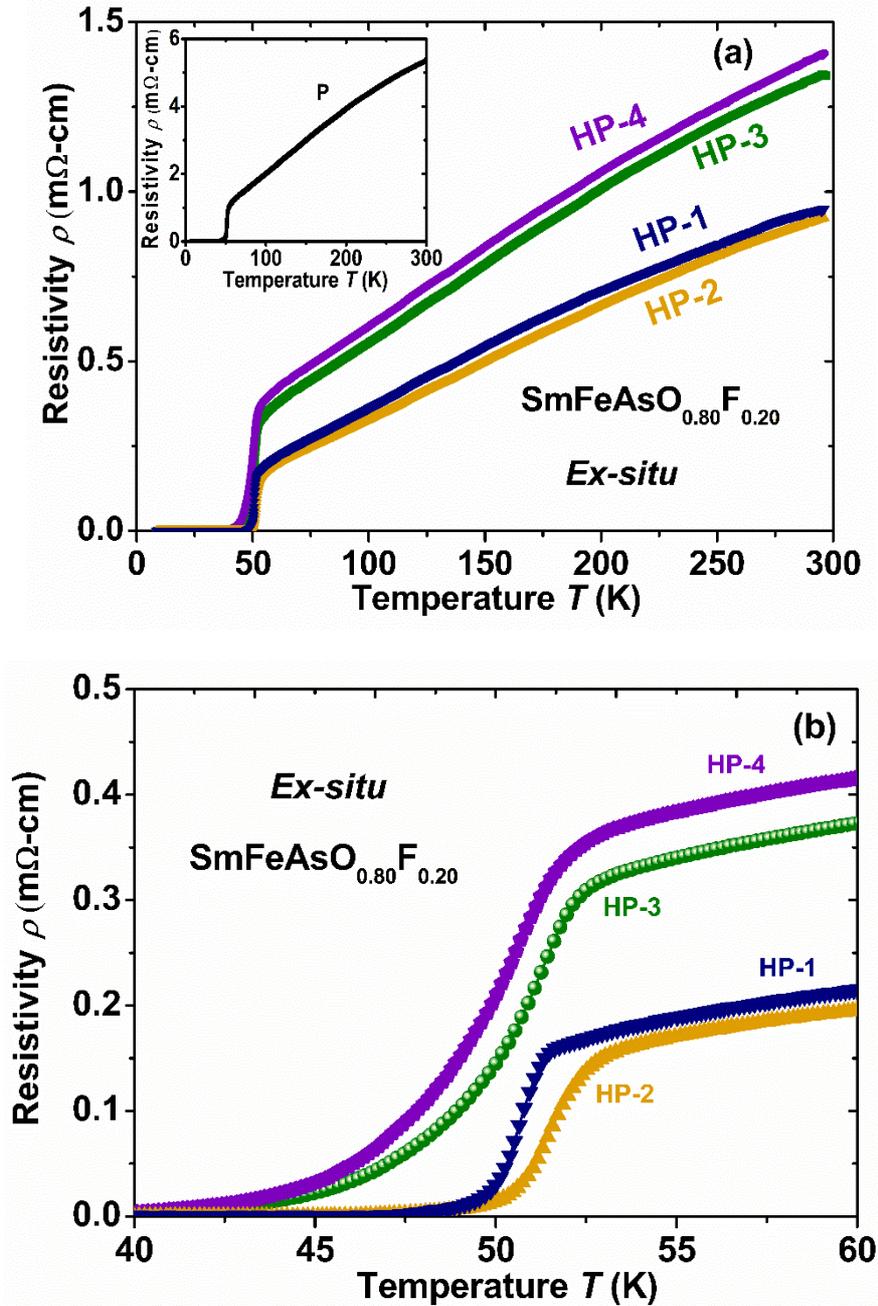

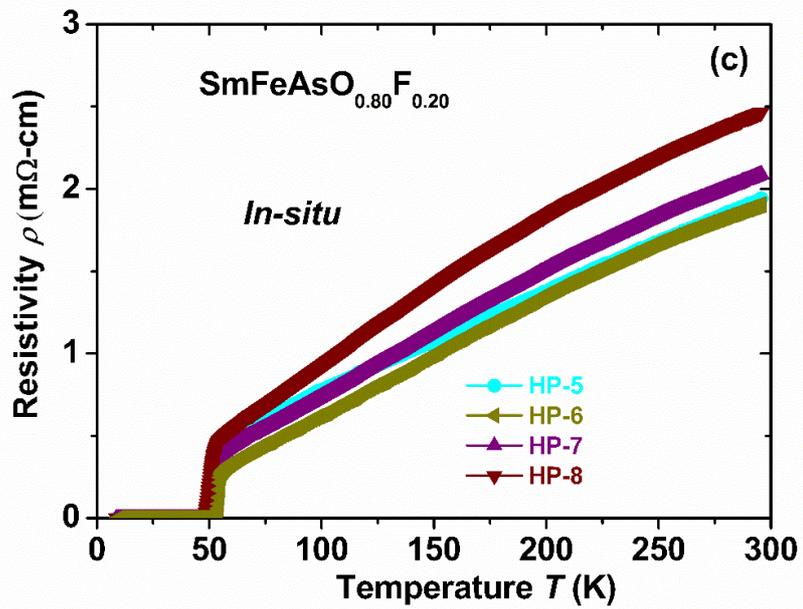

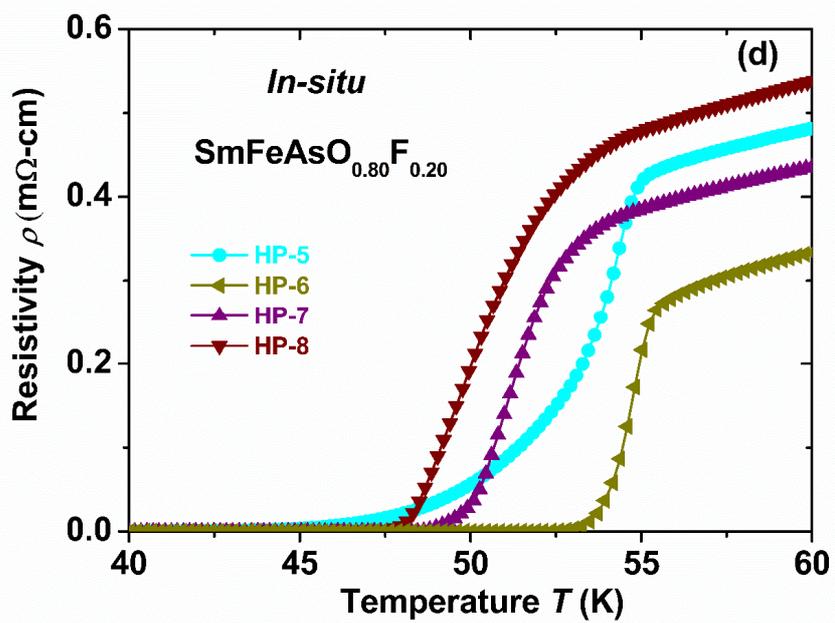

**Figure 7:** The temperature dependence of the normalized magnetic moment in ZFC and FC modes under a magnetic field of 20 Oe for Sm1111 bulks prepared by (**a**) *ex-situ* CA-HP process and (**b**) *in-situ* CA-HP process with the parent P. The variation of the critical current density ($J_c$) at a temperature of 5 K with the applied magnetic field up to 9 T for the parent and Sm1111 bulks prepared by (**c**) the *ex-situ* CA-HP process and (**d**) the *in-situ* CA-HP process.

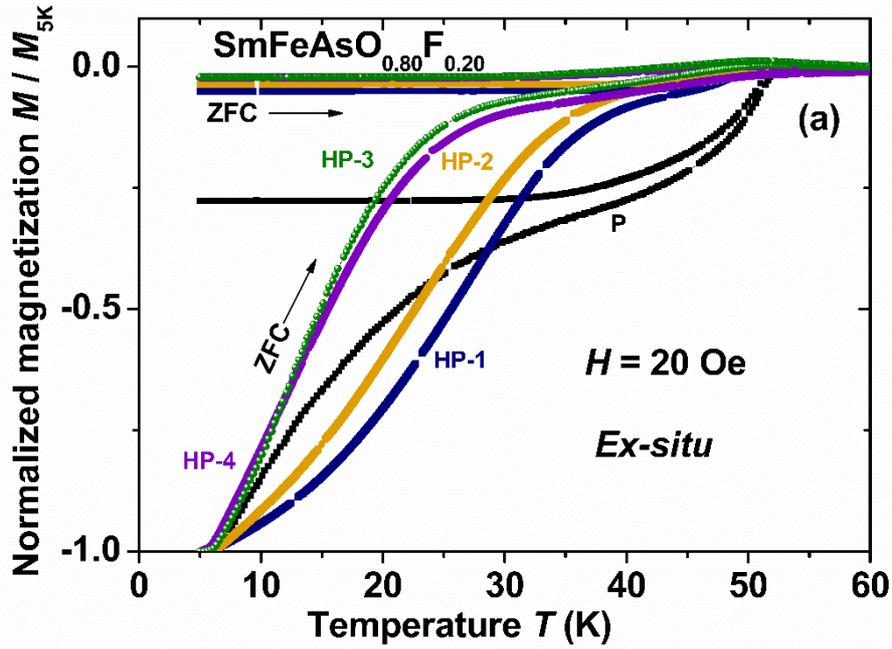

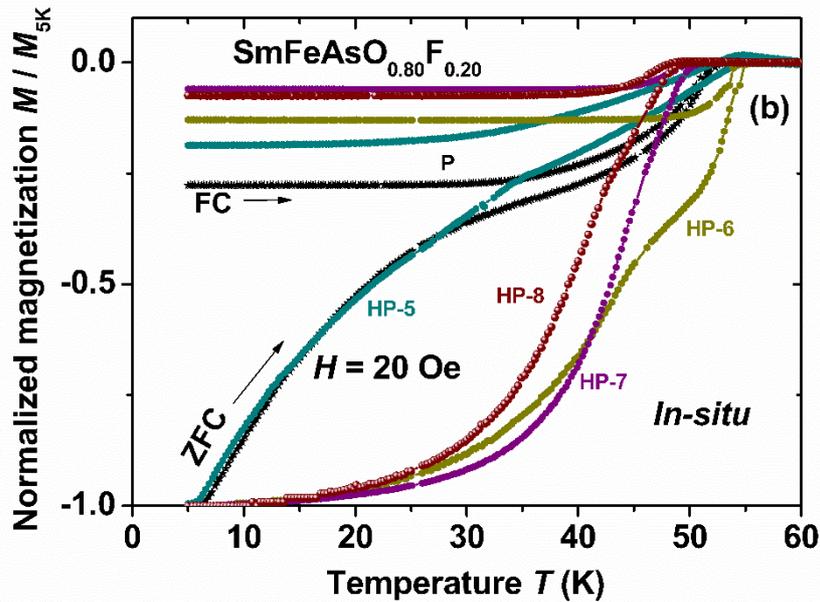

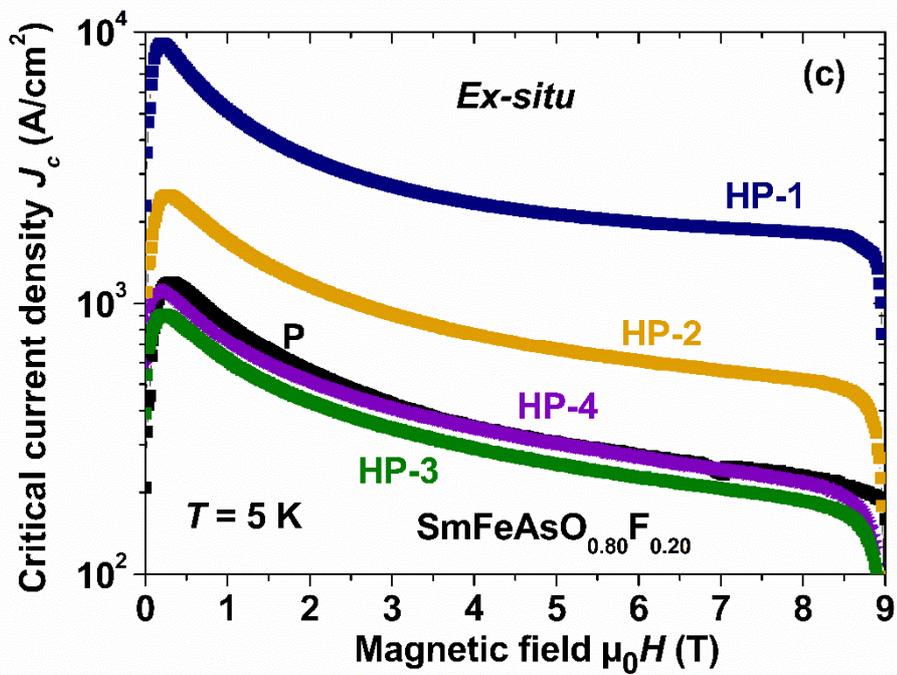

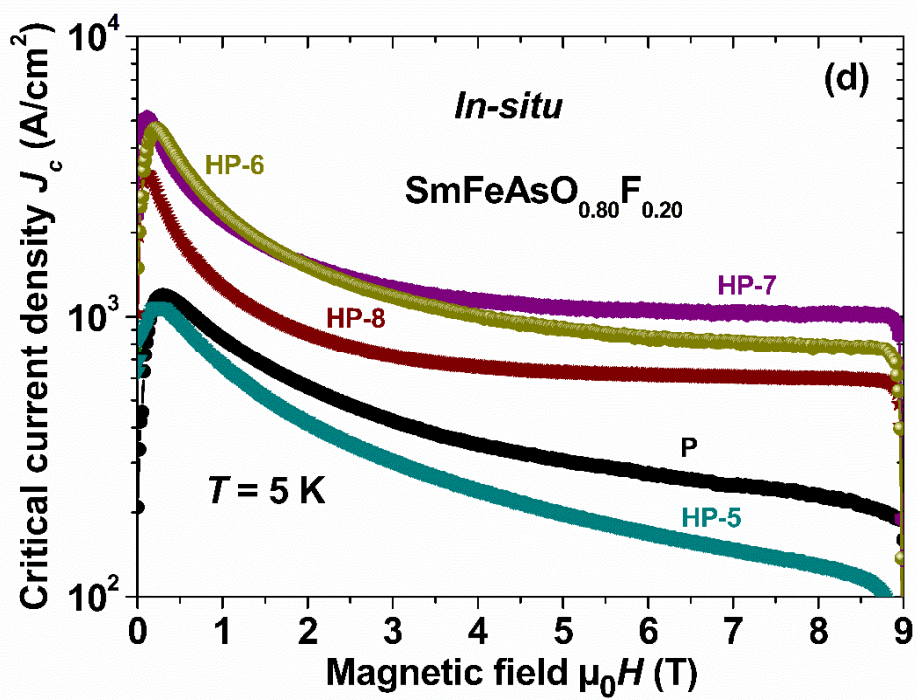

**Figure 8:** The variations of **(a)** the onset transition temperature ($T_c^{onset}$), **(b)** the transition width ($\Delta T$), **(c)** the room temperature resistivity ($\rho_{300\,K}$), **(d)** the Residual Resistivity Ratio ($RRR = \rho_{300\,K} / \rho_{60\,K}$), and **(e)** the critical current density ($J_c$) of different SmFeAsO$_{0.80}$F$_{0.20}$ bulks prepared by the *ex-situ* CA-HP process under the different growth pressures (0.5-4 GPa).

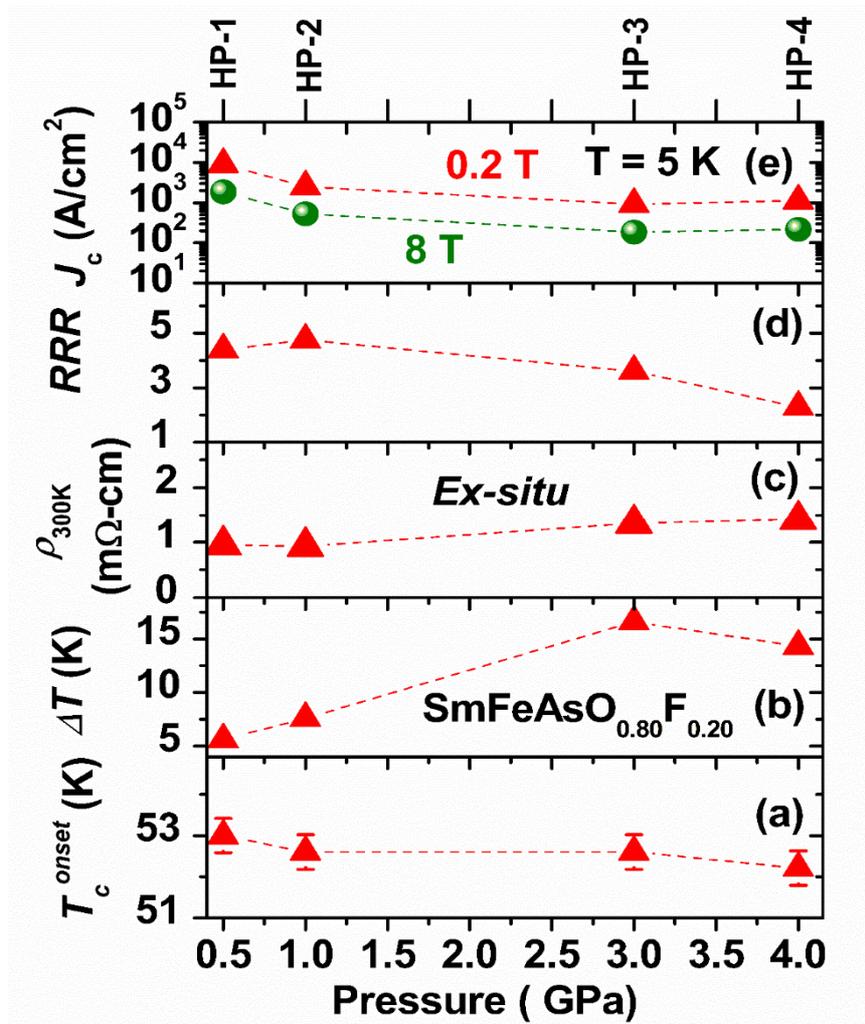

**Figure 9:** The variations of **(a)** the onset transition temperature ($T_c^{onset}$), **(b)** the transition width ($\Delta T$), **(c)** the room temperature resistivity ($\rho_{300\,K}$), **(d)** the Residual Resistivity Ratio ($RRR = \rho_{300\,K} / \rho_{60\,K}$), and **(e)** the critical current density ($J_c$) of different SmFeAsO$_{0.80}$F$_{0.20}$ bulks prepared by the *in-situ* CA-HP process under the different growth temperatures (1100-1600 °C) and times.

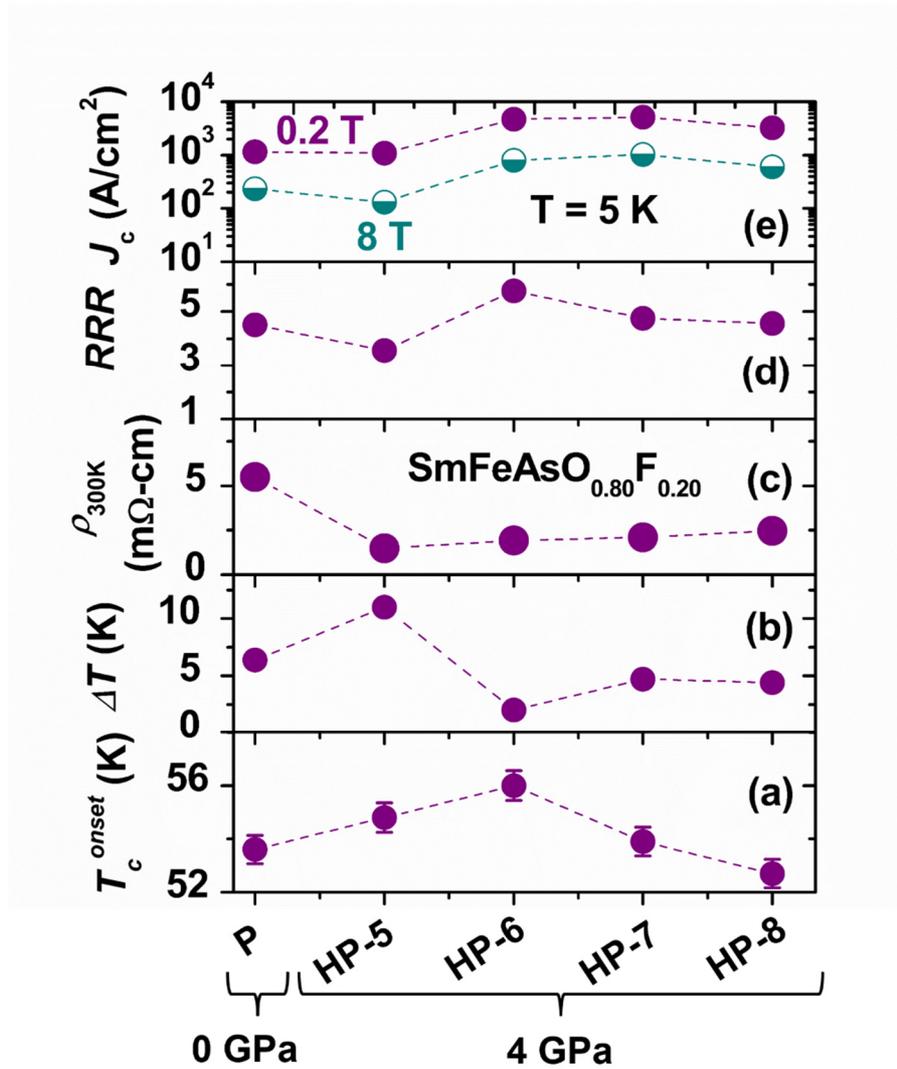

**Figure 10:** Magnetic field-dependent critical current density ($J_c$) for the best sample SmFeAsO$_{0.80}$F$_{0.20}$ prepared by CSP, SPS [40], HP-HTS (HIP) [30], *ex-situ* and *in-situ* CA-HP. Sm1111 was prepared at 45 MPa, 900°C for 5-10 min by SPS [40] and 0.5 GPa, 900 °C for 1 h by HP-HTS methods.

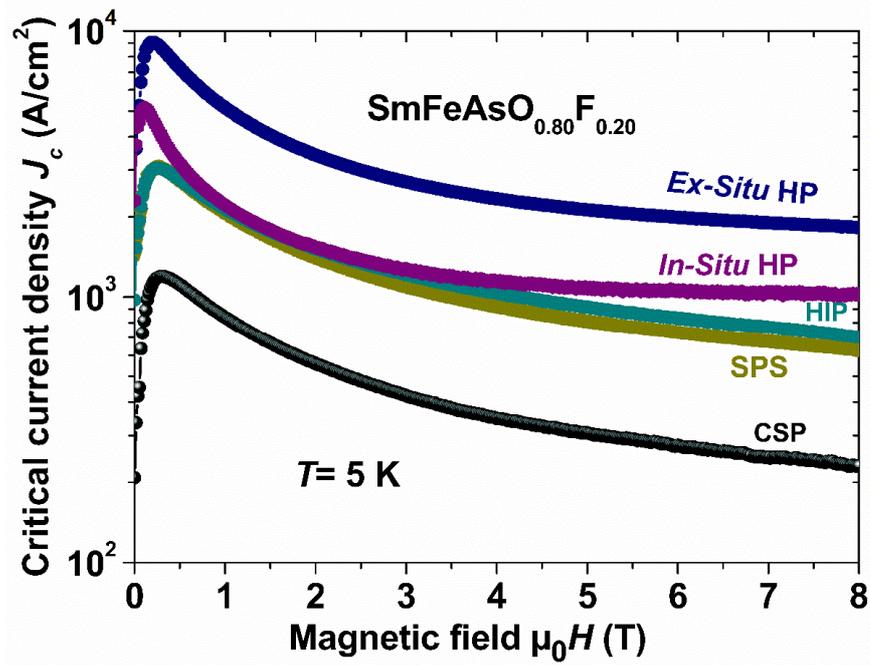

# Optimization of superconducting properties of F-doped SmFeAsO by cubic anvil high-pressure technique

Mohammad Azam[1], Tatiana Zajarniuk[2], Hiraku Ogino[3], Shiv J. Singh[1*]

*[1]Institute of High Pressure Physics (IHPP), Polish Academy of Sciences, Sokołowska 29/37, 01-142 Warsaw, Poland*

*[2]Institute of Physics, Polish Academy of Sciences, Aleja Lotników 32/46, 02-668 Warsaw, Poland*

*[3]Research Institute for Advanced Electronics and Photonics, National Institute of Advanced Industrial Science and Technology (AIST), Tsukuba, Ibaraki 305-8568, Japan*

*Corresponding author:

Email: sjs@unipress.waw.pl

https://orcid.org/0000-0001-5769-1787

**Figure S1:** Elemental mapping of SmFeAsO$_{0.80}$F$_{0.20}$ samples synthesized by CA-HP process using *ex-situ* and *in-situ* processes for: **(i)** HP-2, **(ii)** HP-3, and **(iii)** HP-4. **(iv)** HP-5 **(v)** HP-7 and **(vi)** HP-8. In each figure, the first image shows the SEM microstructure, while the last image presents a composite elemental map combining all the constituent elements, respectively. The intermediate images show individual elemental distributions of Sm (Samarium), Fe (Iron), As (Arsenic), O (Oxygen), and F (Fluorine).

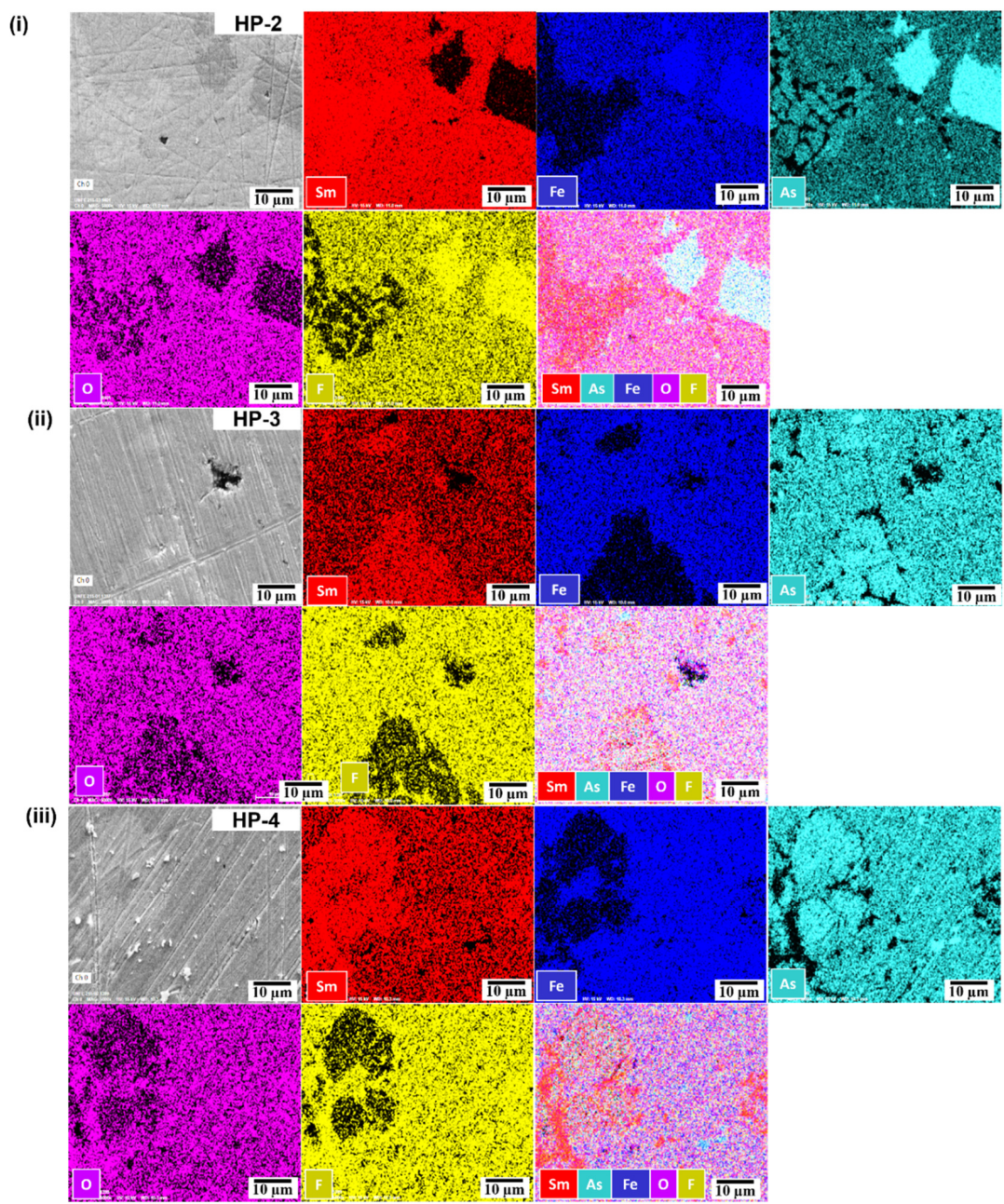

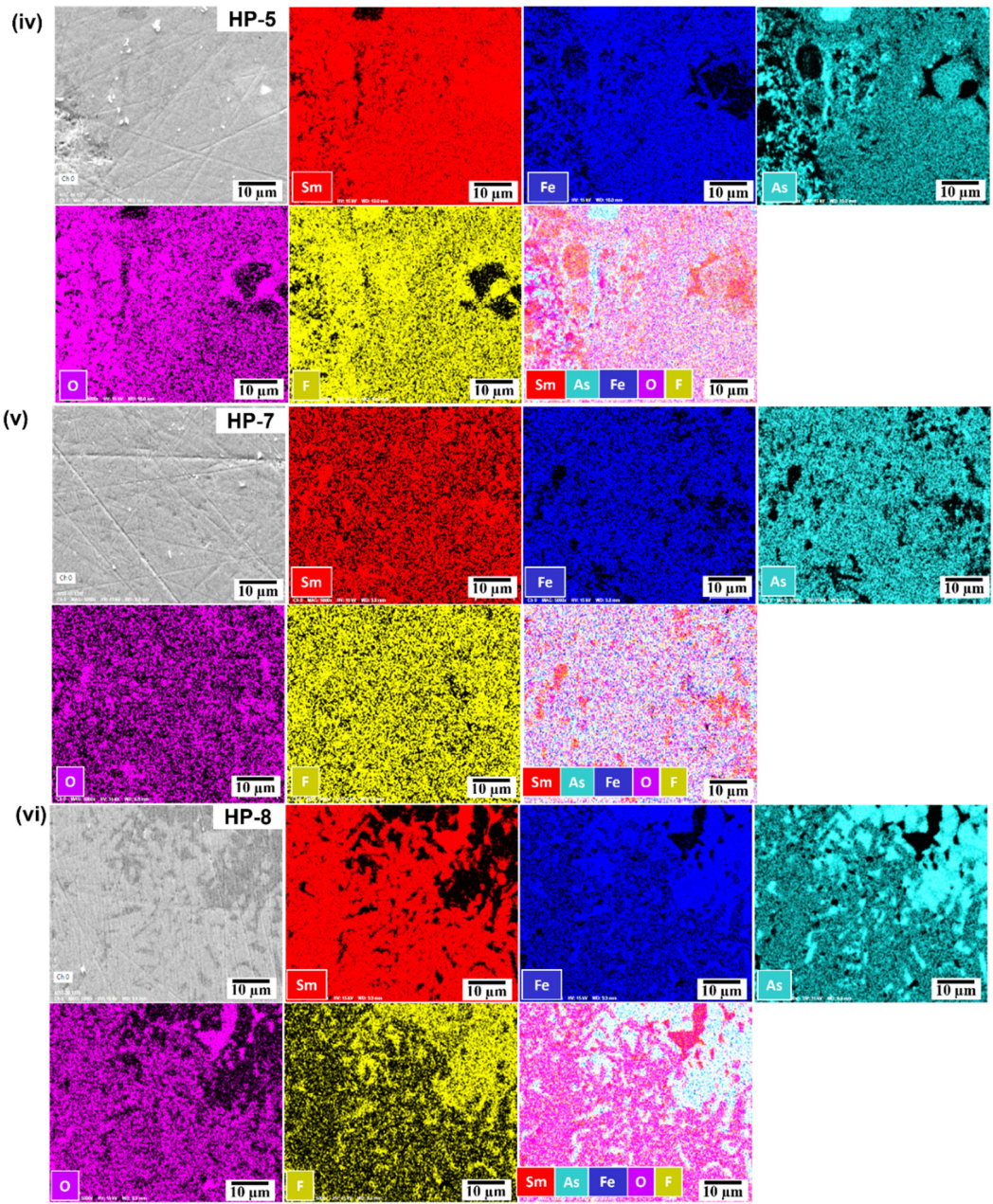

**Figure S2:** Backscattered electron images (BSE; AsB) of SmFeAsO$_{0.80}$F$_{0.20}$ samples prepared by CA-HP process using *ex-situ* and *in-situ* processes for: **(a)-(c)** HP-2, **(d)-(f)** HP-3, and **(g)-(i)** HP-4 **(j)-(l)** HP-5 **(m)-(o)** HP-7 and **(p)-(r)** HP-8. Bright contrast, light grey, and black contrast correspond to Sm$_2$O$_3$ (SmOF), SmFeAsO$_{0.80}$F$_{0.20}$ phase and pores, respectively. One can note that black contrast can occasionally be SmAs/FeAs.

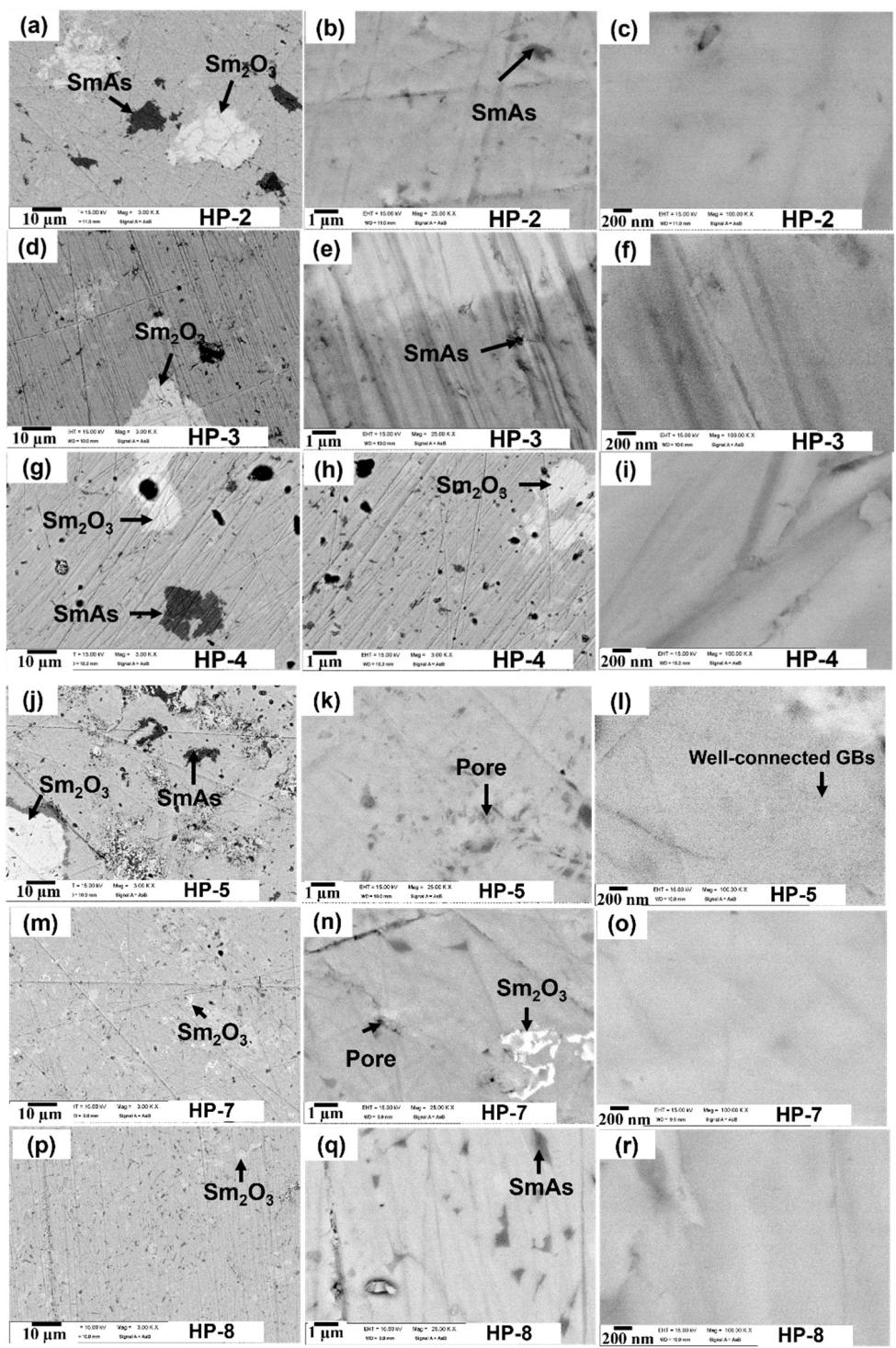

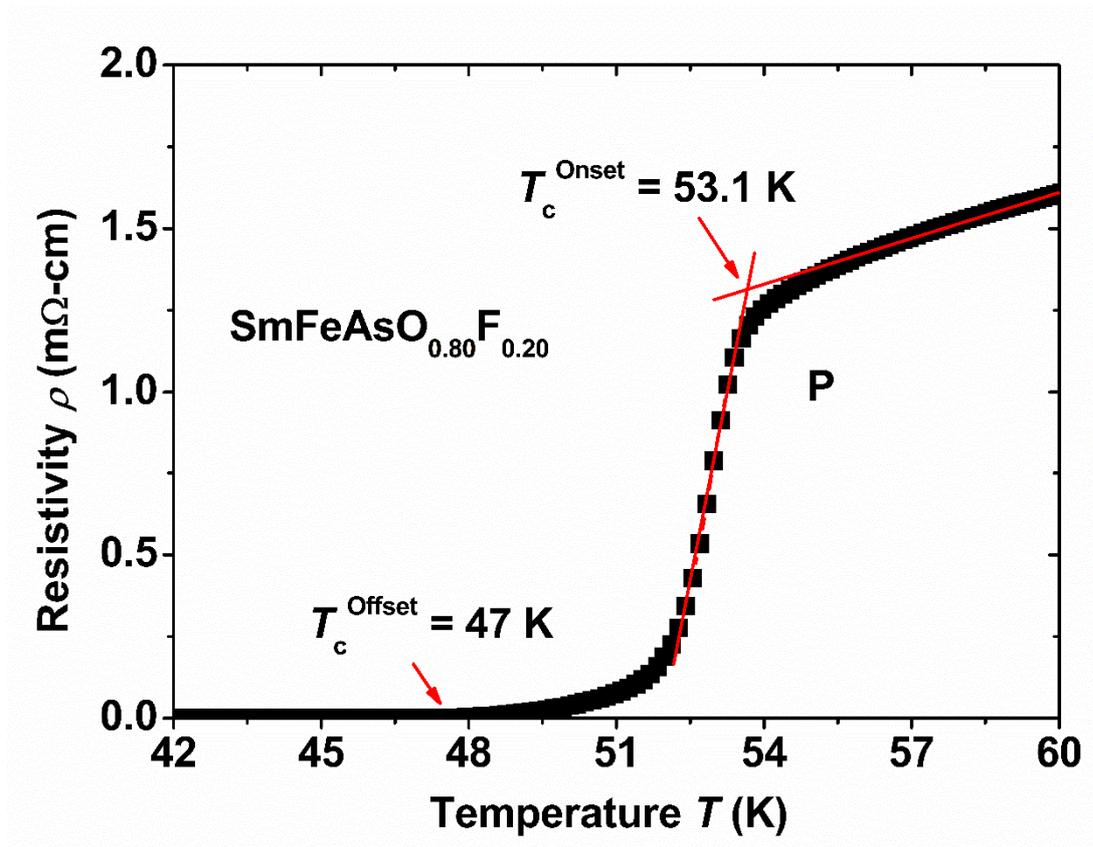

**Figure S3:** The variation of temperature dependence of the resistivity of the parent P sample for low temperatures.

**Figure S4:** The magnetic hysteresis loop (*M-H*) for the Parent P, HP-1 and HP-6 samples prepared by the CSP and CA-HP methods.

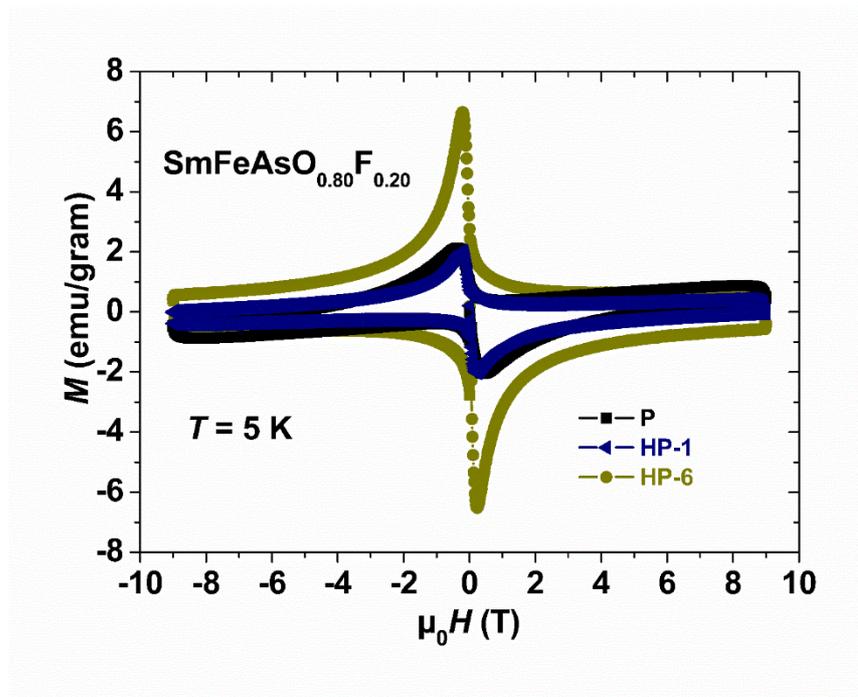

**Figure S5:** The pinning force at 5 K with the applied magnetic field up to 8 T for **(a)** the parent P and *ex-situ* CA-HP processed samples: HP-1, HP-2, HP-3 **(b)** the parent P and *in-situ* CA-HP processed samples: HP-5, HP-6, HP-7, HP-8.

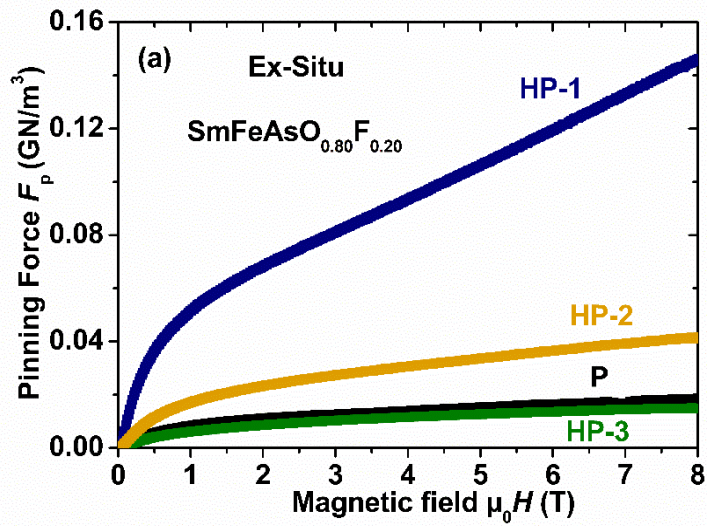

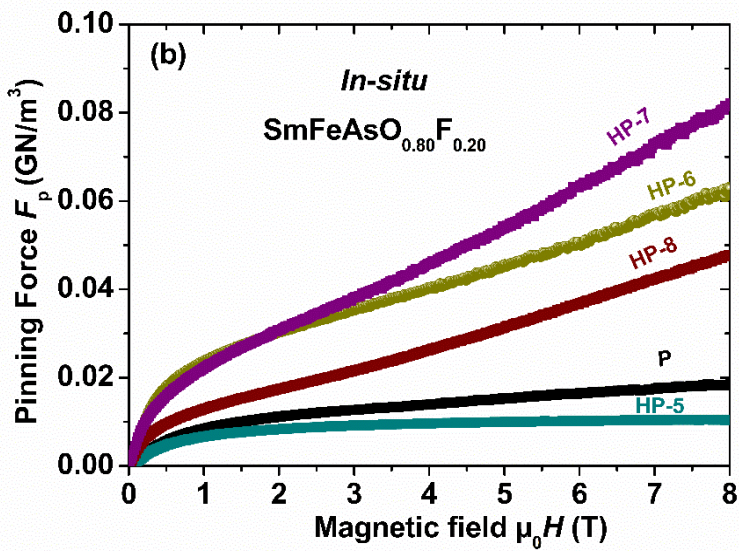